\documentclass[aps, prd, superscriptaddress,
  twocolumn,
  dvips, showpacs, floatfix]{revtex4}
\usepackage{amsmath}
\usepackage{amsfonts}
\usepackage{bm}
\usepackage[final]{graphicx}

\newcommand{\calR}{{\cal R}}

\newcommand{\Dlabel}[1]{\label{#1}}         
\newcommand{\Dcite}[1]{\cite{#1}}   
\newcommand{\Dref}[1]{\ref{#1}}     

\newcommand{\Deqn}[1]{{Eq.\ (\Dref{#1})}}
\newcommand{\Deqns}[1]{{Eqs.\ (\Dref{#1})}}

\newcommand{\be}{\begin{equation}}
\newcommand{\ee}{\end{equation}}

\newcommand{\beq}{\begin{equation}}
\newcommand{\eeq}{\end{equation}}

\newcommand{\bea}{\begin{eqnarray}}
\newcommand{\eea}{\end{eqnarray}}

\newcommand{\lm}{{\ell m}}
\newcommand{\lmm}{{\ell,-m}}
\newcommand{\dR}{{\delta\!R}}

\newcommand{\ret}{{\text{ret}}}
\newcommand{\adv}{{\text{adv}}}
\newcommand{\isco}{{\text{is}}}
\newcommand{\dis}{{\text{dis}}}
\newcommand{\con}{{\text{con}}}
\newcommand{\pr}{{\text{pr}}}

\newcommand{\R}{{\text{R}}}
\newcommand{\s}{{\text{S}}}
\renewcommand{\o}{{\text{o}}}
\newcommand{\psio}{{\psi_{\text{o}}}}
\newcommand{\Psio}{{\Psi_{\text{o}}}}
\newcommand{\ro}{{r_\text{o}}}
\newcommand{\rotwo}{{r^2_\text{o}}}
\newcommand{\rothree}{{r^3_\text{o}}}
\newcommand{\rofour}{{r^4_\text{o}}}
\newcommand{\Omegao}{{\Omega_\text{o}}}

\begin{document}
\title{Scalar field self-force effects on
   orbits about a Schwarzschild black hole}
\author{Luz Maria Diaz-Rivera}
\affiliation{Department of Physics, PO Box 118440, University of Florida,
          Gainesville, FL 32611-8440}
\author{Eirini Messaritaki}
\affiliation{Department of Physics, PO Box 118440, University of Florida,
          Gainesville, FL 32611-8440}
\affiliation{Department of Physics,  PO Box 413, University of
Wisconsin-Milwaukee, Milwaukee, WI 53201}
\author{Bernard F. Whiting}
\affiliation{Department of Physics, PO Box 118440, University of Florida,
          Gainesville, FL 32611-8440}
\author{Steven Detweiler}
\affiliation{Department of Physics, PO Box 118440, University of Florida,
          Gainesville, FL 32611-8440}

\date{October 4, 2004}

\begin{abstract}

For a particle of mass $\mu$ and scalar charge $q$, we compute the effects
of the scalar field self-force upon circular orbits, upon slightly
eccentric orbits and upon the innermost stable circular orbit of a
Schwarzschild black hole of mass $m$. For circular orbits the self force
is outward and causes the angular frequency at a given radius to decrease.
For slightly eccentric orbits the self force decreases the rate of the
precession of the orbit. The effect of the self force moves the radius of
the innermost stable circular orbit inward by $0.122701 \times q^2/\mu$,
and it increases the angular frequency of the ISCO by the fraction $
0.0291657 \times q^2/\mu m$.

\end{abstract}
 \maketitle

\section{Introduction}
 \label{Intro}

We consider a small mass $\mu$ with a scalar charge $q$ which orbits a
Schwarzschild black hole. The interaction of the particle with its own
field yields the self force on the particle. We use a perturbation
analysis to find the effects of the scalar-field self-force upon circular
orbits, upon slightly eccentric orbits and upon the location of the
innermost stable circular orbit, the ISCO, of the Schwarzschild geometry.
We view this project as a warm-up for the more interesting gravitational
problem which must deal with more complicated field equations and gauge
issues \Dcite{barack-ori:01, sago-etal:03, NakanoSagoSasaki03}.

In general relativity, a particle of infinitesimal mass moves through
spacetime along a geodesic. If the particle has a small but finite mass
$\mu$ then its worldline deviates from a geodesic of the background
spacetime by an amount proportional to $\mu$. This deviation is said to
result from the ``self force'' of the particle's own gravitational field
acting upon itself.

Newtonian gravity presents an elementary example of a self-force effect
\Dcite{DetPoisson04}. A small mass $\mu$ in a circular orbit of radius
$\ro$ about a more massive companion $m$ has an angular frequency
$\Omegao$ given by
\beq
  \Omegao^2 = \frac{m}{\rothree(1+\mu/m)^2}.
\Dlabel{Kepler1}
\eeq
When $\mu$ is infinitesimal, the large mass $m$ does not move, the radius
of the orbit $\ro$ is equal to the separation between the masses and
$\Omegao^2=m/\ro^3$. However when $\mu$ is finite but still small, both
masses orbit their common center of mass with a separation of
$\ro(1+\mu/m)$, and the angular frequency is as given in \Deqn{Kepler1}.
The finite $\mu$ influences the motion of $m$, which influences the
gravitational field within which $\mu$ moves. This back action of $\mu$
upon its own motion is the hallmark of a self force, and the $\mu$
dependence of \Deqn{Kepler1} is properly described as a Newtonian self
force effect.

A thorough understanding of gravitational waves detected by the Laser
Interferometer Space Antenna \footnote{The LISA web site is located
at http://lisa.jpl.nasa.gov/ .} will require clear theoretical
predictions of possible gravitational wave forms which result from a
small stellar-mass object orbiting a jumbo sized black hole; these
wave forms must include self-force effects.

In studying the gravitational self force, one considers the
particle's gravitational field to be a small perturbation $h_{ab}$ of
the background metric $g_{ab}$. For an object of very small size, the
motion ought to be independent of the particle's structure, and one
is inclined to take the limit of a point particle. However, in that
limit, $h_{ab}$ diverges precisely at the particle, and the concept
of the self force might appear to be ill-defined.

Dirac \Dcite{Dirac38} studied the electromagnetic version of this
problem in flat spacetime and discovered that the part of the actual,
retarded electromagnetic field which is singular and yet exerts no
force on the particle itself is, in a local approximation, the
Coulomb field and could be identified as the average of the retarded
and advanced electromagnetic fields.  The remainder, half of the
difference between the retarded and advanced fields, is a vacuum
solution of Maxwell's equations and accounts entirely for the
particle's self force.

The self force includes the radiative reaction force or radiation
reaction\Dcite{Jackson3rd}.
 For a particle with
electrical charge $q$ and acceleration $\vec a$, the Abraham-Lorentz
force describes the response of a particle to its own radiation and
is proportional to $q^2 d\vec a/dt$. The factor of $q^2$ results from
the charge $q$ interacting with its own electric field, which is also
proportional to $q$. Similarly the gravitational radiation reaction
force on a small mass $\mu$ is proportional to $\mu^2$. Other parts
of the self force are not directly related to radiation but are
properly described as the particle interacting with its own field and
are also proportional to $q^2$ or $\mu^2$.

A curved-spacetime generalization of Dirac's approach is now
available \Dcite{DeWittBrehme60, Mino97, QuinnWald97, DetWhiting03,
poisson:03}. For gravitation, an expansion about the position of the
particle describes the singular ``S'' part of the field
$h^\text{S}_{ab}$ which exerts no force on the particle and is a
local solution of the perturbed Einstein equations with the particle
as the source.
 The remainder ``R'' part of the field $h^\text{R}_{ab}  = h_{ab} -
h^\text{S}_{ab}$ is, locally, a source-free solution of the perturbed
Einstein equations, with the combined metric $g_{ab} +
h^\text{R}_{ab}$ being a vacuum solution of the Einstein equations
through first order in $h^\text{R}_{ab}$.
 The effect of the self force has the particle moving along a geodesic of
the vacuum geometry $g_{ab}+h^\text{R}_{ab}$.

A caveat remains: in curved spacetime the ``S'' and ``R'' fields
cannot be described in terms of the advanced and retarded fields.
However, the mode-sum regularization procedure pioneered by Barack
and Ori \Dcite{BarackOri00, Barack00, Lousto00, BarackOri02,
BarackOri03, BMNOS02, Mino02, DetMessWhiting03, Messaritaki03,
DongHoon04} is of use for background geometries amenable to the
decomposition of fields in terms of scalar, vector and tensor
harmonics. This procedure has been applied to self-force calculations
for the Schwarzschild geometry involving both scalar
\Dcite{BarackBurko00, Burko00, DetMessWhiting03} and gravitational
\Dcite{BarackLousto02} fields. In general terms, the ``S'' field is
singular at the particle, but each of its spherical harmonic
components is finite; these components are the ``regularization
parameters''. The regular ``R'' field is determined by subtracting
each $\ell,m$ mode of the ``S'' field, from the corresponding
$\ell,m$ mode of the actual retarded field. The sum over modes of the
difference between the retarded and ``S'' fields provides the ``R''
field which governs the self-force effects on the particle.

In this manuscript we treat the self force from the scalar field
$\Psi$ in a perturbative manner.  At zeroth order in $\Psi$, a
geodesic for the particle is chosen and this determines the
particle's singular field $\Psi^\s$ in the neighborhood of the
geodesic. Then the actual scalar field $\Psi$, with appropriate
boundary conditions, is found everywhere. The difference of the
actual field and the singular ``S'' field provides the regular
remainder $\Psi^\R=\Psi-\Psi^\s$. Finally we determine the first
order in $\Psi$ self-force effects of the scalar field $\Psi^\R$
acting back on the particle and changing the worldline away from the
original geodesic. The small effects of $\Psi^\R$ appear as an
acceleration of the worldline of order $q^2/\mu m^2$. We determine
the effects of the scalar field self-force upon the angular
frequency, energy and angular momentum of a general circular orbit,
upon the rate of precession of a slightly eccentric orbit and upon
the location and angular frequency of the innermost stable circular
orbit.

In Section \Dref{overview} we give an overview of the self force and
how it affects the location of the ISCO.

In Section \Dref{worldlines} we describe how a generic scalar field
affects the worldline of a particle in the Schwarzschild geometry.

In Section  \Dref{circularorbits} we consider scalar field self-force
effects on circular orbits, and numerically calculate the
corresponding changes of energy, angular momentum and angular
frequency. In particular the self-force effect on the right hand side
of \Deqn{Kepler}, below, is derived. We also give the results of a
numerical computation of $\Psi^\R$ evaluated at the location of the
particle, which changes the effective inertial mass $\mu$ of the
particle.

In Section \Dref{precession} we consider the scalar-field self-force
effects on slightly eccentric orbits of the Schwarzschild geometry.
Specifically, we find the self-force correction to the rate of the
precession of the orbit and to the right hand side of \Deqn{Mercury},
below. In Section \Dref{isco}, this analysis is applied to an orbit
near $r=6m$ to obtain the self force effect upon the radius and
angular frequency of the ISCO.

The discussion of Section \Dref{discussion} distinguishes between the
scalar and gravitational self force effects. We do not attempt to
generalize our limited results to the case of the gravitational
self-force.

In Appendix \Dref{sourcedecomp} we give some details of the spherical
harmonic decomposition of a source moving along a slightly eccentric
orbit. These details are required for the numerical calculation of
the actual field for a slightly eccentric orbit.

In Appendix \Dref{regularizationparameters} we describe the
regularization parameters which are required for the self-force
analysis of slightly eccentric circular orbits of the Schwarzschild
geometry. These are used in the computation of $\Psi^\R$ described in
Sections \Dref{circularorbits}--\Dref{isco}.

We use Schwarzschild coordinates $(t,r,\theta,\phi)$ on the spacetime
manifold. The position of the particle in these coordinates is
$(T,R,\Theta,\Phi)$, and $s$ measures the proper time along a
worldline. For a general worldline, the angular frequency is
$\Omega_\phi\equiv d\Phi/dT$. The subscript ${}_\o$ is reserved
exclusively for quantities related to circular orbits: for a circular
orbit, the radius and angular frequency are $\ro$ and $\Omegao$,
respectively, while $\Psi_\o$ is the scalar field from a particle in
a circular orbit. The mass and charge of a particle are $\mu$ and
$q$, and the mass of the black hole is $m$. The scalar field is
$\Psi$; however, we define $\psi\equiv q\Psi/\mu$ as a combination
which occurs often in the description of the effects of a scalar
field on the motion of a particle. In this perturbative analysis we
always assume that $q^2/\mu \ll m$.

In equations concerning the self force, expressions containing $\psi$
refer to the regularized field and must be evaluated at the location
of the particle.

\section{Conceptual framework}
\Dlabel{overview}

\subsection{Dissipative and conservative forces}

Mino \Dcite{Mino03} examines the self force on a particle in orbit
about a rotating black hole with a focus on how the Carter constant
\Dcite{Carter68a} evolves. Following Mino's ideas, one is naturally
led to divide the self force into two parts which depend upon how
each part changes under a change in the boundary conditions. One part
is ``dissipative'' and usually associated with radiation reaction.
The other part is ``conservative''.  This distinction is most easily
described in terms of ``Green's functions'' which yield the parts of
the field responsible for the dissipative and conservative parts of
the self force.

Let $G^\ret$ be the retarded Green's function, which provides the
actual, physical field $\Psi^\ret$ for the problem of interest, and
let $G^\s$ be the Green's function \Dcite{DetWhiting03} for the
singular part of the field $\Psi^\s$. The regularized field $\Psi^\R
= \Psi^\ret - \Psi^\s$ provides the complete self force. The change
in boundary conditions from outgoing radiation at infinity to
incoming radiation is effected by using the advanced Green's function
$G^\adv$ and its field $\Psi^\adv$, rather than the retarded
quantities.

The dissipative part of the self force changes sign under the
interchange of the retarded and advanced Green's functions. Thus,
 the dissipative part of the field $\Psi^\dis$ is uniquely determined by
using a ``Green's function''
\beq
  G^\dis = \frac12 G^\ret - \frac12 G^\adv.
\eeq
$\Psi^\dis$ is a source-free solution of the field equation and is
regular at the particle; no use of $\Psi^\s$ is required to find the
dissipative part of the self force.

The conservative part of the self force is invariant under the
interchange of the retarded and advanced Green's functions. The
conservative part of the field represents the half advanced plus half
retarded field, but this is singular at the particle and requires
regularization by the removal of $\Psi^\s$. Thus, the conservative
part of the regularized field is uniquely determined near the
particle by using a ``Green's function''
\beq
  G^\con = \frac12 G^\ret + \frac12 G^\adv - G^\s.
\eeq
In a neighborhood of the particle, the resulting field is a source free
solution of the field equation and is regular.

A circular orbit about a Schwarzschild black hole provides an example
which clearly distinguishes the dissipative from the conservative
parts of the self force. For a particle in a circular orbit about a
black hole, $\partial_t\Psi$ and $\partial_\phi\Psi$ are both
dissipative, as can be understood in terms of time-reversal
invariance. We see below in \Deqns{Edot} and (\Dref{Jdot}) that
$\partial_t\Psi$ and $\partial_\phi\Psi$ are responsible for removing
energy and angular momentum, respectively, from the particle at a
rate which precisely balances the loss of energy and angular momentum
outward through a distant boundary and into the black hole. With
$t\rightarrow-t$ and $\phi\rightarrow-\phi$ the motion of the
particle is nearly identical, only the boundary conditions on the
scalar field are changed, and now energy and angular momentum
entering the system through the boundary are deposited on the
particle. The dissipative self-force is small in this perturbative
analysis but its effect accumulates over time as the particle spirals
slowly inward or outward, depending upon the boundary conditions.

For a circular orbit $\partial_r\Psi$ is conservative and, as shown
below, provides a small addition to the centripetal force which
affects the angular frequency via a change in the right hand side of
\Deqn{Kepler}. However, the angular frequency is unchanged under
$t\rightarrow-t$ and $\phi\rightarrow-\phi$ which changes the
direction of radiation flow at the boundary. $\partial_r\Psi$ and its
effect on the angular frequency is independent of the direction of
radiation imposed by the boundary conditions. With either boundary
condition the conservative self-force effect on the frequency is
small but the effect on the phase of the orbit accumulates over time.

If a particle is not in a circular orbit, then these simple
relationships between the components of $\partial_a\psi$ and the
conservative and dissipative parts of the self force no longer hold.

\subsection{Stability of a circular orbit}

The notion of the stability of a circular orbit in the context of the
self-force warrants further discussion.

A stability problem presupposes the existence of a mechanical system
in equilibrium. A perturbation analysis of the system's small
oscillations often reveals that the natural frequencies of the system
are complex eigenvalues dependent upon some set of parameters. The
sign of the imaginary part of a natural frequency determines whether
the amplitude of a small oscillation grows, diminishes or stays
constant. The simplest cases are when the frequency is either purely
real or purely imaginary. An imaginary frequency with an appropriate
sign then signifies an unstable mode.

For the system of a small particle of mass $\mu$ orbiting a larger black
hole of mass $m$, the particle moves along a geodesic, if $\mu$ is
considered infinitesimal.  In this case it is well known that for a
circular orbit at Schwarzschild coordinate $r=\ro$ the angular frequency
$\Omegao$, with respect to Schwarzschild coordinate time, is given by
\beq
  \Omegao^2=\frac{m}{\ro^3} .
\Dlabel{Kepler}
\eeq
If the orbit has a small eccentricity then elementary analysis \Dcite{MTW}
reveals that the frequency $\Omega_r$ of the radial oscillations is
determined by
\beq
 \Omega_r^2 = \frac{m}{\ro^4}(\ro-6m) .
\Dlabel{Mercury}
\eeq
If $\ro$ is very large, then $\Omegao$ and $\Omega_r$ are nearly
equal, and the slightly eccentric orbit is an ellipse obeying
Kepler's laws. The difference between $\Omegao$ and $\Omega_r$ leads
to the rate of the precession of the ellipse,
\beq
  \Omega_\pr = \Omegao - \Omega_r,
\eeq
and is directly responsible for the relativistic contribution to the
precession of the perihelion of Mercury. For an orbit just inside
$6m$, the two solutions for $\Omega_r$ in \Deqn{Mercury} are both
imaginary and one corresponds to an unstable mode. The orbit at
$\ro=6m$ is the innermost stable circular orbit---the ISCO.

For realistic boundary conditions, the expectation is that at a large
separation the emission of gravitational radiation from a binary
circularizes the orbit \Dcite{ApostolatosKennefickOriPoisson93} and
causes the two objects to spiral slowly towards each other.

When $\ro$ is outside but comparable to $6m$, gravitational radiation
evolves the orbit slowly inward on a secular timescale $m/\mu$ times
longer than the dynamical timescale, with the rate of inspiral
$d\ro/dt = O(\mu/m)$. The particle makes a transition to a plunge
\Dcite{Ori00b} when it reaches the ISCO, and the plunge occurs over a
dynamical timescale. The resulting gravitational wave form appears as
a sinusoid with a slowly increasing frequency until the angular
frequency of the ISCO is reached. Then, after a brief burst from the
plunge, the wave form is determined by the frequencies of the most
weakly damped free oscillations of the black hole
\Dcite{ChandraDet75, Det79, Det80a}.

The secular evolution caused by gravitational radiation keeps the
particle from ever being in true equilibrium, and the question of the
stability of a circular orbit may seem ill posed. Nevertheless, the
gravitational wave form changes character at a frequency near that in
\Deqn{Kepler} for $\ro=6m$ even though the particle is never actually
in equilibrium. The dependence of the transition frequency upon the
scalar field self force is a major focus of this manuscript.

To form a well-posed problem related to the ISCO and the angular
frequency of the transition to a plunge, we consider an unphysical
system with boundary conditions which have equal amounts of outgoing
and incoming radiation. Such a system must be constructed by use of a
regularized half-advanced and half-retarded Green's function. In this
case, the dissipative force vanishes entirely while the conservative
force and its effects upon $\Omegao$ and $\Omega_r$ are unchanged
from the case with realistic outgoing-wave boundary conditions. With
equal amounts of outgoing and incoming radiation, equilibrium
configurations exist, and stability analyses may proceed.

In Section V we show that the conservative force from $\nabla_a\Psi$
changes the right hand side of \Deqn{Mercury}, and therefore the
location of the actual ISCO by a fractional amount of order $q^2/\mu
m$ for a scalar field; for gravity this fractional amount would be of
order $\mu/m$.

\section{Scalar field effects on a particle's motion}
\Dlabel{worldlines}

\subsection{Description of motion}
The functions $\left(T(s), R(s), \pi/2, \Phi(s)\right)$ describe a
particle's worldline in the equatorial plane of a black hole in
Schwarzschild coordinates $(t,r,\theta,\phi)$. The proper time $s$ is
measured along the worldline, and the four velocity
\beq
  u^a = d x^a/ ds
\eeq
is normalized to unity, $u^a u_a = -1$. Additionally, the worldline
and the Killing vector fields $t^a \partial/\partial x^a \equiv
\partial/\partial t$ and $\phi^a
\partial/\partial x^a \equiv \partial/\partial\phi$ define
\beq
  E \equiv - t^a u_a \qquad\text{and}\qquad   J \equiv \phi^a u_a,
\eeq
which, for geodesic motion, are the energy and angular momentum, per
unit rest mass, respectively.
 The components of $u^a$ are thus
\beq
  u^a = \left(\frac{E}{1-2m/R}, \dot{R}, 0, J/R^2 \right),
\Dlabel{uAeqn}
\eeq
where the dot denotes a derivative with respect to $s$. For any
worldline the angular frequency, with respect to Schwarzschild
coordinate time, is
\beq
 \Omega_\phi \equiv \dot\Phi/\dot T = \frac{d\Phi}{dT}
        = \frac{J(R-2m)}{E R^3}.
\Dlabel{omegaeqn}
\eeq
 The normalization of $u^a$ implies that
\beq
  u^a u_a = -1 = - \frac{(E^2-\dot{R}^2)}{1-2m/R} + \frac{J^2}{R^2}
\eeq
or
\beq
  E^2 - \dot{R}^2 = (1-2m/R)(1+J^2/R^2) .
\Dlabel{normeqn}
\eeq

\subsection{Scalar field modifications of the geodesic equation}
\Dlabel{geodesics}

Quinn \Dcite{Quinn00} considers the interaction of a scalar field
$\Psi$ with a particle of constant bare mass $\mu_0$ and constant
scalar charge $q$. He carefully demonstrates that the equation of
motion is
\beq
  u^b\nabla_b(\mu u_a) = \dot\mu u_a
                + \mu u^b \nabla_b u_a = q \nabla_a \Psi,
\Dlabel{Quinn}
\eeq
where $\mu = \mu_0 - q\Psi$. This result can be obtained from a
general action principle.  In what follows, $\Psi$ is assumed to be
the regular field $\Psi^\R$ evaluated at the particle. Thus, the
quantity $\mu$ changes when the value of the scalar field at the
particle changes. The fractional change in $\mu$ is
\beq
  \Delta\mu/\mu_0 = - q\Psi/\mu_0,
\Dlabel{restmass}
\eeq
which is shown in Figs.~1 and 2 for circular orbits of the
Schwarzschild geometry.

The projection of \Deqn{Quinn} orthogonal to $u^a$ yields the
acceleration from the self-force
\beq
  u^b \nabla_b u_a = (q/\mu) (g_a{}^b + u_a u^b) \nabla_b \Psi
\Dlabel{eom}
\eeq
which modifies the worldline of the particle through spacetime. In
our perturbative analysis, we only consider the first order effects
of the scalar field. The right hand side of \Deqn{eom} is explicitly
first order in $\Psi$, and the change in $\mu$ has only a second
order effect on the acceleration. Thus, for the purposes of
describing the world line of the particle we treat $\mu$ as constant,
and it is convenient to define $\psi \equiv q\Psi/\mu$. For
simplicity we also assume that the scalar field is symmetric under
reflection through the equatorial plane and that the particle remains
in the equatorial plane.

The components of \Deqn{eom} yield
\beq
  dE/ds = \dot{E} = -\partial_t \psi + E u^b\nabla_b \psi
\Dlabel{Edot}
\eeq
\beq
  dJ/ds = \dot{J}=  \partial_\phi \psi + J u^b\nabla_b \psi
\Dlabel{Jdot}
\eeq
and
\bea
  \ddot{R} &=& -\frac{m(E^2-\dot{R}^2)}{R(R-2m)}
                             + \frac{R-2m}{R^4} J^2
\nonumber\\ && {} +
  \frac{R-2m}{R} \partial_r \psi + \dot{R} u^b\nabla_b\psi .
\Dlabel{rddoteqn}
\eea

The normalization (\Dref{normeqn}) is a first integral of the
equation of motion, and \Deqn{rddoteqn} follows directly from
\Deqns{normeqn}, (\Dref{Edot}) and (\Dref{Jdot}), which form a
complete set of equations describing the equatorial motion of a
particle interacting with a scalar field.

Together, \Deqns{normeqn} and (\Dref{rddoteqn}) imply that
\bea
  \ddot{R} &=& -\frac{m}{R^2} + \frac{R-3m}{R^4} J^2
\nonumber\\ && {}
   + \frac{R-2m}{R} \partial_r \psi +  \dot{R} u^b\nabla_b \psi ,
\Dlabel{rddot}
\eea
which is convenient for analyzing slightly eccentric orbits.

For a particle in circular motion, $\partial_r\psi$ is independent of
the direction of radiation at the boundaries and provides a
conservative force.  However, $\partial_t\psi$ and
$\partial_\phi\psi$ on the right hand sides of \Deqns{Edot} and
(\Dref{Jdot}) are dissipative components of the self force and change
sign if the direction of radiation at the boundary is changed.  In
fact, the conservation of energy and angular momentum guarantee that
\Deqns{Edot} and (\Dref{Jdot}) are consistent with the flow of energy
and angular momentum across the boundaries. This is manifest by
matching appropriate independent solutions of the source-free wave
equation at the orbit of the particle. If a particle is not in a
circular orbit, then these simple relationships between the
components of $\partial_a\psi$ and the conservative and dissipative
parts of the self force no longer hold.

\section{Self force effects on Circular orbits}
\Dlabel{circularorbits}

 \begin{figure}
 \includegraphics[clip,angle=0,scale=0.32]{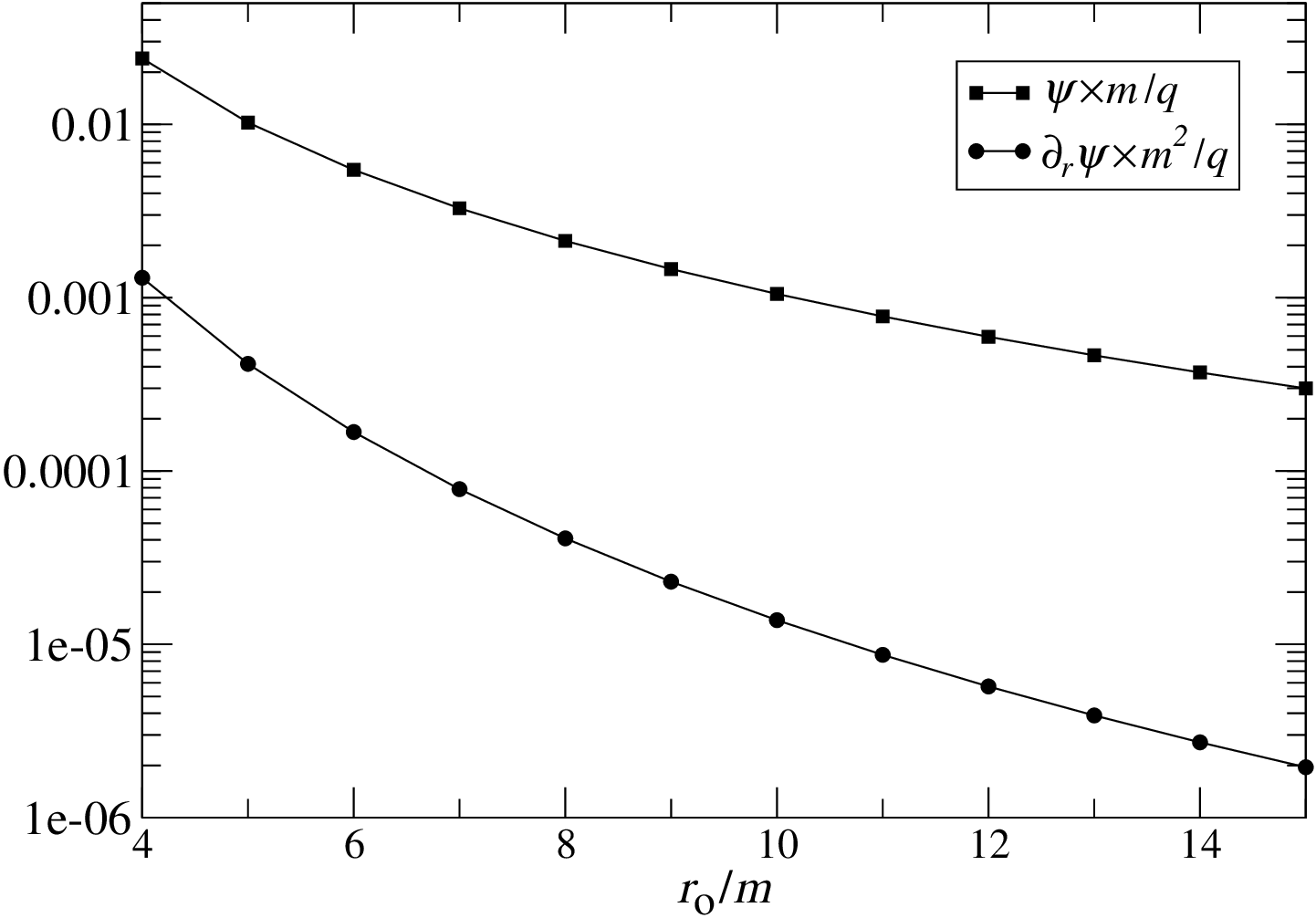}
 \caption{\label{dpdrFigA} The regularized field $\Psio^\R m/q$
 at the particle and the
 radial component of the self force $(\partial_r\Psio^R)m^2/q$ as a
 function of the radius $\ro$ for circular orbits close to the black hole.}
 \end{figure}

 \begin{figure}
 \includegraphics[clip,angle=0,scale=0.32]{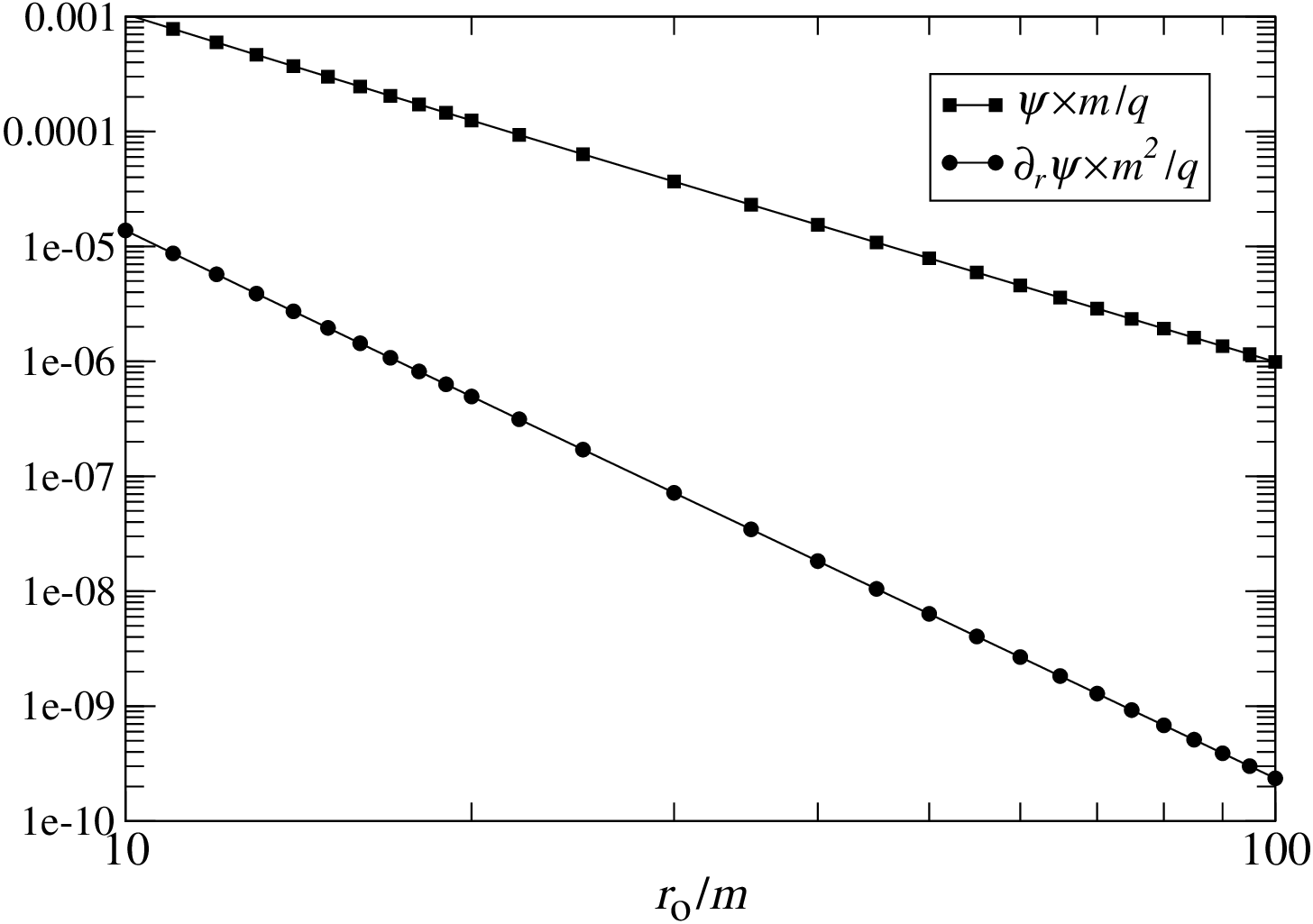}
 \caption{\label{dpdrFigB} The same as Fig. \Dref{dpdrFigA}
 except for larger radii.}
 \end{figure}

The scalar field $\psio$ of a charged particle in a circular orbit
rotates with the particle and has a symmetry described by
\beq
 \partial_t \psio + \Omegao \partial_\phi \psio = 0.
\Dlabel{LiePsi}
\eeq
In particular, this implies that $u^a\nabla_a \psio = 0$ at the
particle and simplifies the description of the self force for
circular motion. A worldline is ``instantaneously circular'' at
radius $\ro$ if $\dot{R}=\ddot{R}=0$. In this case
\beq
  J^2 = \frac{\rofour}{\ro-3m}
      \left(\frac{m}{\rotwo} - \frac{\ro-2m}{\ro} \partial_r \psio \right)
\Dlabel{J2eqn}
\eeq
follows from \Deqn{rddot}, and
\beq
  E^2 = \frac{(\ro-2m)^2}{\ro(\ro-3m)}
      \left(1 - \ro \partial_r \psio \right) 
\Dlabel{E2eqn}
\eeq
then follows from \Deqn{normeqn}.
The angular frequency is given by \Deqn{omegaeqn}
\beq
  \Omegao^2 =
  \frac{m}{\rothree} - \frac{\ro-3m}{\rotwo} \partial_r \psio + O(\psi^2).
\Dlabel{Omega2}
\eeq
These reduce to the usual expressions for the circular geodesics of
the Schwarzschild geometry when the scalar field $\psio$ is removed.
The resulting fractional changes in $J$, $E$ and $\Omegao$ caused by
the self force for an orbit at radius $\ro$ are
\beq
 \frac{\Delta J}{J} = - \frac{\ro(\ro-2m)}{2m} \partial_r \psio ,
\Dlabel{DeltaJ}
\eeq
\beq
  \frac{\Delta E}{E} = -\frac 12 \ro \partial_r \psio ,
\Dlabel{DeltaE}
\eeq
and
\beq
  \frac{\Delta\Omegao}{\Omegao}
           = - \frac{\ro(\ro-3m)}{2m} \partial_r \psio .
\Dlabel{DeltaOm}
\eeq

 \begin{table*}
 \begin{ruledtabular}
 \begin{tabular}{clllll}
   $\ro/m$ & \quad $\Psio^\R m/q$\ \ \ %
                & $\partial_r\Psio^\R m^2/q$
                & $F_r\mu m^3/q^2$ & $F_t\mu m^2/q^2$ & $F_\phi\mu m/q^2$ \\
  \hline
   4 &   --0.02398775  & 0.001302375  \\
   5 &   --0.01023418  & 4.149937e--4 \\
   6 &   --0.005454828 & 1.6772834e--4 & --1.3813756e--4& 0.00492388  & --0.0282217   \\
   7 & --0.003275343   & 7.850679e--5  & --6.10826e--5  & 0.00245981  & --0.0175331   \\
   8 & --0.002127506   & 4.082502e--5  & --2.91665e--5  & 0.00135506  & --0.0118991   \\
  10 & --0.001049793   & 1.378448e--5  & --8.21672e--6  & 5.07681e--4 & --0.00642025  \\
  14 & --3.700646e--4  & 2.720083e--6  & --1.19280e--6  & 1.19675e--4 & --0.00264001  \\
  20 & --1.246728e--4  & 4.937906e--7  & --1.55296e--7  & 2.68798e--5 & --0.00105972  \\
  30 & --3.661710e--5  & 7.171924e--8  & --1.54904e--8  & 5.08927e--6 & --3.82435e--4 \\
  50 & --7.889518e--6  & 6.346791e--9  & --8.53532e--10 & 6.44498e--7 & --1.07242e--4 \\
  70 & --2.877222e--6  & 1.284529e--9  & --1.26087e--10 & 1.66932e--7 & --4.65496e--5 \\
  100& --9.884245e--7  & 2.356504e--10 & --1.65135e--11 & 4.00531e--8 & --1.92277e--5 \\
  200& --1.239866e--7  & 8.642538e--12 & --3.11790e--13 & 2.51466e--9 & --3.44412e--6
 \Dlabel{Circletable}
 \end{tabular}
 \end{ruledtabular}
 \caption{A selection of our computed values of the regularized scalar
 field evaluated at the particle and the radial component of the self
 force, for circular orbits of radius $\ro$. Also shown are $F_r$, $F_t$
 and $F_\phi$, defined in \Deqns{Ft}-(\Dref{Fr}), for slightly eccentric
 orbits with $\ro>6m$; for $\ro=6m$, their limiting values are given; for $\ro<6m$
 slightly eccentric orbits do not exist and $F_r$, $F_t$
 and $F_\phi$ are not defined. }
 \end{table*}

 \begin{figure}
 \includegraphics[clip, angle=0,scale=0.32]{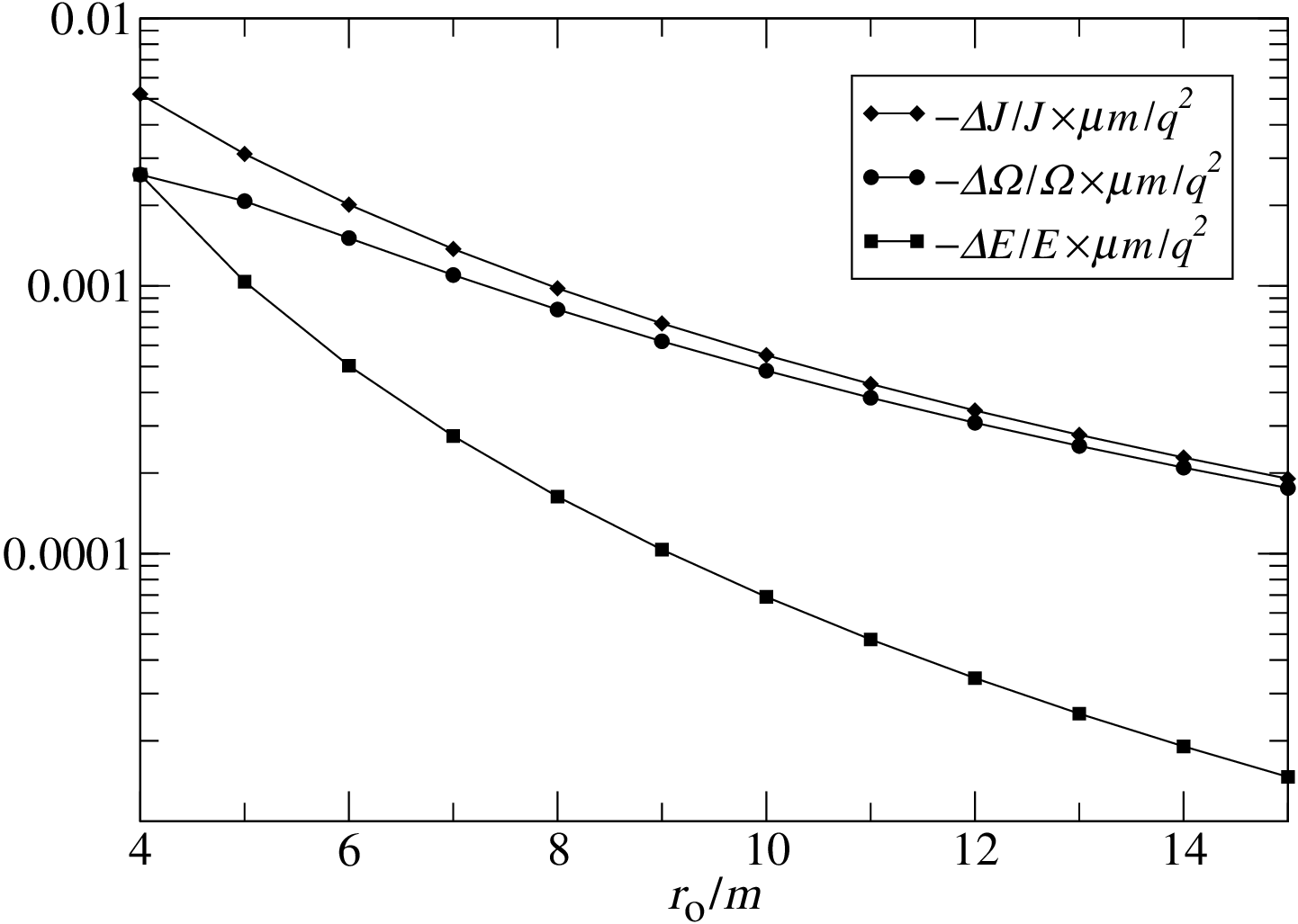}
\caption{\label{JEOFigA} The fractional changes in $J$, $E$ and
$\Omegao$, from the self force as a function of the radius $\ro$ for
circular orbits close to the black hole. These quantities are
calculated using \Deqns{DeltaJ}--(\Dref{DeltaOm}).}
 \end{figure}

 \begin{figure}
\includegraphics[clip,angle=0,scale=0.32]{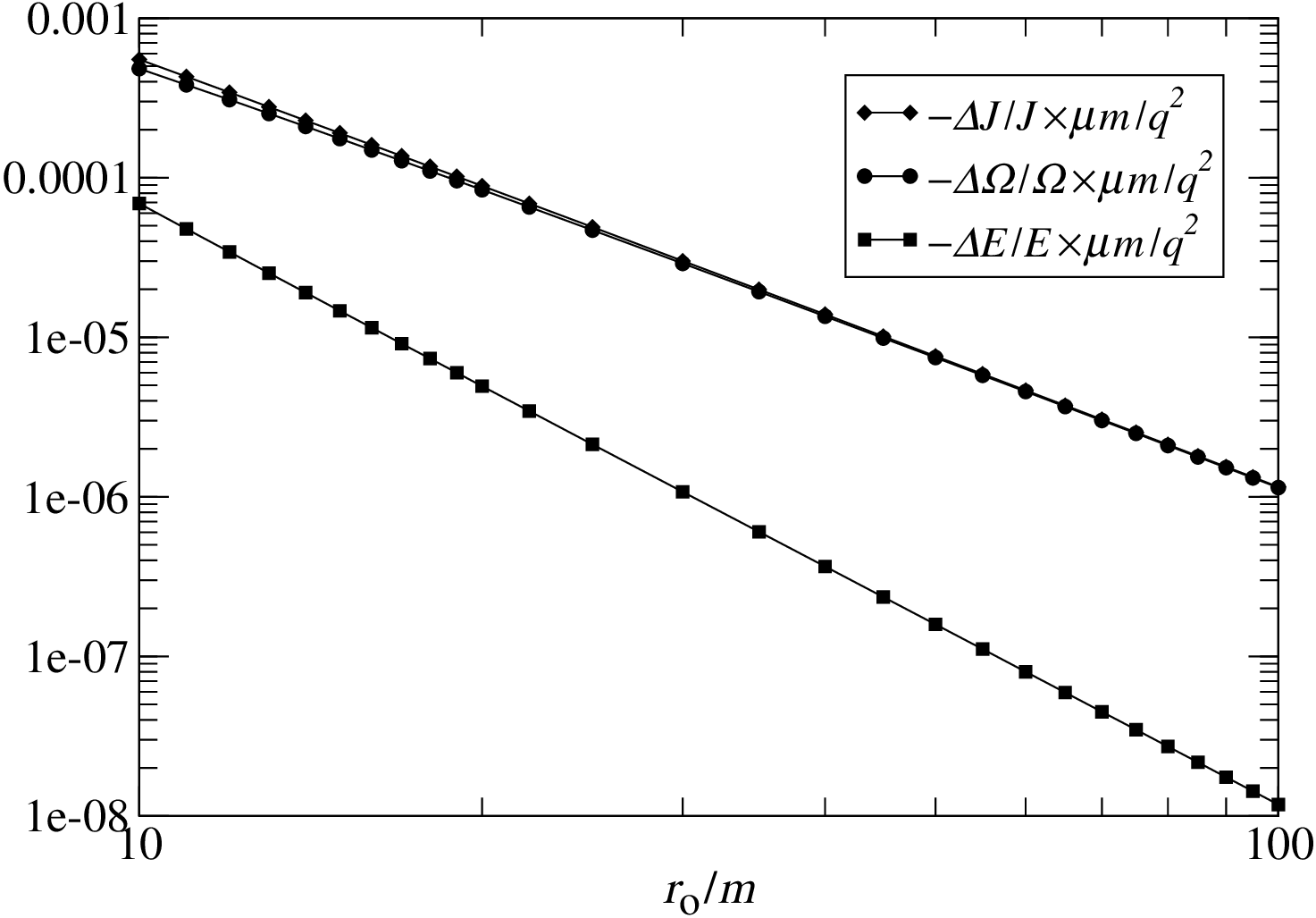}
\caption{\label{JEOFigB} The same as Fig. \Dref{JEOFigA} except for
larger radii.}
 \end{figure}

Refs.~\Dcite{Breuer73, DetMessWhiting03, Burko00} give detailed
descriptions of the calculation of the retarded scalar field for a
particle in a circular orbit of the Schwarzschild geometry.
Refs.~\Dcite{ BarackOri00, BarackOri02, BMNOS02, Mino02,
DetMessWhiting03, Messaritaki03, DongHoon04} describe the evaluation
of the required regularization parameters for the self force. In
Appendix \Dref{regularizationparameters}, we give the regularization
parameters for the scalar field $\psi^\R$ from
Ref.~\Dcite{Messaritaki03}. Refs.~\Dcite{ Burko00, DetMessWhiting03}
describe implementations of the Barack and Ori regularization
procedure for the numerical evaluation of the self force for circular
orbits of the Schwarzschild geometry.

Figs.~\Dref{dpdrFigA} and \Dref{dpdrFigB} show the regularized
$\psi^\R$ and the radial component of the self force
$\partial_r\psi^\R$ evaluated at the particle as functions of the
radius of the orbit. Figs.~\Dref{JEOFigA} and \Dref{JEOFigB} show the
fractional changes in $J$, $E$ and $\Omegao$ from the radial
component of the self force. Table 1 gives a selection of our
numerical calculations of both $\psi^\R$ and $\partial_r\psi^\R$ at
the particle as a function of $R$. The force is outward, falls off
approximately as $r_{\text{o}}^{-5}$, as noted by
Burko\Dcite{Burko00}, and $\psi^\R$ falls off as $\ro^{-3}$ for large
$\ro$. Each of $\Omegao$, $E$ and $J$ for a circular orbit is
decreased by the self force.

\section{Self force effects on slightly eccentric orbits}
 \Dlabel{precession}

A particle in an orbit with small eccentricity, in the Schwarzschild
geometry, has periodic motion in the radial direction with a
frequency $\Omega_r$ given in \Deqn{Mercury}. In this section we
describe the effects of the self force on $\Omega_r$ and upon the
rate of precession of the orbit $\Omega_\pr$. We require that the
scalar field $\psi$ have boundary conditions with equal amplitudes of
incoming and outgoing radiation at both the event horizon and
infinity. This simplifies the discussion of $\Omega_r$ and allows us
to pose a well defined stability problem for the ISCO. The
conservative ``Green's function'' $G^\con$ determines the regularized
field.

\subsection{Slightly eccentric geodesics}

In this perturbative analysis, a slightly eccentric geodesic about a
fixed radius $\ro$ with a small amplitude, $\dR \ll q^2/\mu \ll\ro
q^2/\mu^2 \ll\ro$, is described by
\beq
  R(T(s))
       =  \ro + \dR \cos(\Omega_r T(s))
\Dlabel{R(t)a}
\eeq and
\beq
  \Phi(T) = \Omegao T + \frac{d \Omega_\phi}{dR}
               \frac{\dR}{\Omega_r} \sin (\Omega_r T).
\Dlabel{Phi(t)a}
\eeq
The remainder of the analysis treats $\dR$ as a small quantity, and
only terms through first order in $\dR$ are retained. The angular
frequency of this geodesic orbit is a function of time,
\beq
  \Omega_\phi \equiv \frac{d\Phi}{dT}
         = \Omegao + \frac{d\Omega_\phi}{dR} \dR \cos(\Omega_r T) .
\eeq
The frequencies $\Omegao$ and $\Omega_r$ are given in \Deqns{Kepler} and
(\Dref{Mercury}). The quantity $d\Omega_\phi/dR$ represents the change in
angular frequency, from \Deqn{omegaeqn}, with respect to a change in
radius while $E$ and $J$ are held constant,
\beq
  \frac{d\Omega_\phi}{dR} = -\frac{2(R-3m)}{R(R-2m)} \Omega_\phi .
\eeq
The radial velocity is
\beq
  \frac{dR}{dT} = - \Omega_r \dR \sin(\Omega_r T) .
\Dlabel{dRdt}
\eeq

\subsection{Self force effect upon J for slightly eccentric orbits}

For slightly eccentric orbits, with self force effects included, $J$ is
not a constant of the motion, even with no dissipation.  The effect of the
self force on $J$ is given in \Deqn{Jdot},
\bea
 dJ/ds &=& \partial_\phi \psi + J u^a\nabla_a \psi \quad \text{ or}
\nonumber\\
  dJ/dT  &=& (u^t)^{-1}\partial_\phi \psi
 + J \left(\partial_t + \Omega_\phi \partial_\phi \right) \psi
\nonumber\\ &&
 + J \frac{dR}{dT} \partial_r\psi .
\Dlabel{Jdota}
\eea

The $\psi$ for a slightly eccentric orbit is described in Appendix
\Dref{sourcedecomp}. We note here that it has two parts. The larger
part $\psio$ is equal to the field which would result from pure
circular motion at $\ro$ and consists of frequencies which are
integral multiples of $\Omega_\o$. For a circular orbit, in general,
$\left(\partial_t + \Omegao
\partial_\phi \right)\psio = 0$, and with conservative
boundary conditions $\partial_t \psio = \partial_\phi \psio = 0$ at
the particle.  The smaller part is proportional to $\dR$. In Appendix
\Dref{sourcedecomp} we show that for slightly eccentric orbits an
expansion for $\psi$ around a circular orbit at $\ro$ gives
\beq
  \partial_t \psi = - F_t \Omega_r \dR\sin(\Omega_r T) ,
\Dlabel{Ft}
\eeq
\beq
  \partial_\phi \psi = - F_\phi \Omega_r \dR\sin(\Omega_r T),
\Dlabel{Fphi}
\eeq
and
\beq
  \partial_r \psi = [\partial_r \psio]_\ro + F_r \dR\cos(\Omega_r T),
\Dlabel{Fr}
\eeq
where $F_t$, $F_\phi$ and $F_r$ depend only upon $\ro$ and, are
independent of both $\dR$ and $t$. In this section the subscript
$\ro$ on  $[\partial_r \psio]_\ro$ implies that $\partial_r \psio$ is
to be evaluated at the circular orbit $\ro$ and not at the actual
location of the particle.
 In \Deqn{Jdota} the coefficient of $\partial_r\psi$ is already first order
in $\dR$, and it is sufficient to use only the circular orbit value
$[\partial_r\psio]_\ro$ at the particle. Thus,
\bea
  \frac{dJ}{dR} &=& \frac{dJ}{dT} \left(\frac{dR}{dT}\right)^{-1}
\nonumber\\  &=&
   J F_t + E F_\phi
        + J [\partial_r\psio]_\ro \Dlabel{dJdR} ,
\eea
where we have used \Deqns{dRdt}, (\Dref{Ft}) and (\Dref{Fphi}), along with
\Deqns{uAeqn}-(\Dref{normeqn}).

\subsection{Self force effects upon $\Omega_r$ for
slightly eccentric orbits}

The exact radial equation of motion (\Dref{rddot}), which includes the
self-force, is
\bea
  \ddot{R} &=& -\frac{m}{R^2} + \frac{R-3m}{R^4} J^2
\nonumber\\ && {}
       + \frac{R-2m}{R} \partial_r \psi + \dot{R} u^b \nabla_b \psi .
\Dlabel{Rddot}
\eea
For describing slightly eccentric orbits, we expand this equation
around the circular orbit at $\ro$ by letting $R\rightarrow \ro +
\dR\cos(\Omega_r T)$ and dropping all terms of order $\dR^2$. The
$O(\dR^0)$ part is equivalent to \Deqn{J2eqn} for the circular orbit
value of $J^2$. The $\dot{R} u^b \nabla_b \psi$ term is $O(\dR^2)$.
The part of \Deqn{Rddot} which is first order in $\dR$ is
\bea
  \ddot{R} &=& \frac{d}{dR}
  \left[ -\frac{m}{R^2} + \frac{R-3m}{R^4} J^2 \right.
\nonumber\\ && \left. \quad
  {}+ \frac{R-2m}{R} \partial_r \psi\right]_\ro \dR
       \cos(\Omega_rT).
\Dlabel{dRddot}
\eea
After use of \Deqn{Fr}, which gives $d(\partial_r \psi)/dR = F_r$, this
becomes
\bea
  \ddot{R}
    &=& \left[ -\frac{m(\ro-6m)}{\rothree(\ro-3m)}
               + \frac{3(\ro-4m)(\ro-2m)}{\rotwo(\ro-3m)}
               [\partial_r\psio]_\ro\right.
\nonumber\\ && {}
               + \frac{2m}{\rotwo} [\partial_r\psio]_\ro
               + \frac{(\ro-3m)}{\rofour} \frac{dJ^2}{dR}
\nonumber\\ && {} \left.
               + \frac{(\ro-2m)}{\ro} F_r \right] \dR \cos(\Omega_rT).
\Dlabel{Omegar}
\eea
\Deqn{R(t)a} implies that
\beq
  \ddot{R} = -\frac{\Omega_r^2 E^2  \dR \cos(\Omega_rT)}{(1-2m/\ro)^2} .
\eeq
Using \Deqn{E2eqn} for $E^2$, we finally obtain
\bea
  \Omega_r^2
    &=& \frac{m}{\ro^{4}}(\ro-6m)
     - \frac{2J}{\ro^{5}} (\ro-3m)^2 (J F_t + EF_\phi)
\nonumber\\ &&
      - \frac{1}{\ro^{3}}(\ro-2m)(\ro-3m)
         (3[\partial_r\psio]_\ro + r F_r)
\Dlabel{Omegar2}
\eea
through first order in $\psi$. This provides the scalar-field self-force
correction to \Deqn{Mercury} for $\Omega_r^2$.

 \begin{figure}
 \includegraphics[clip,angle=0,scale=0.32]{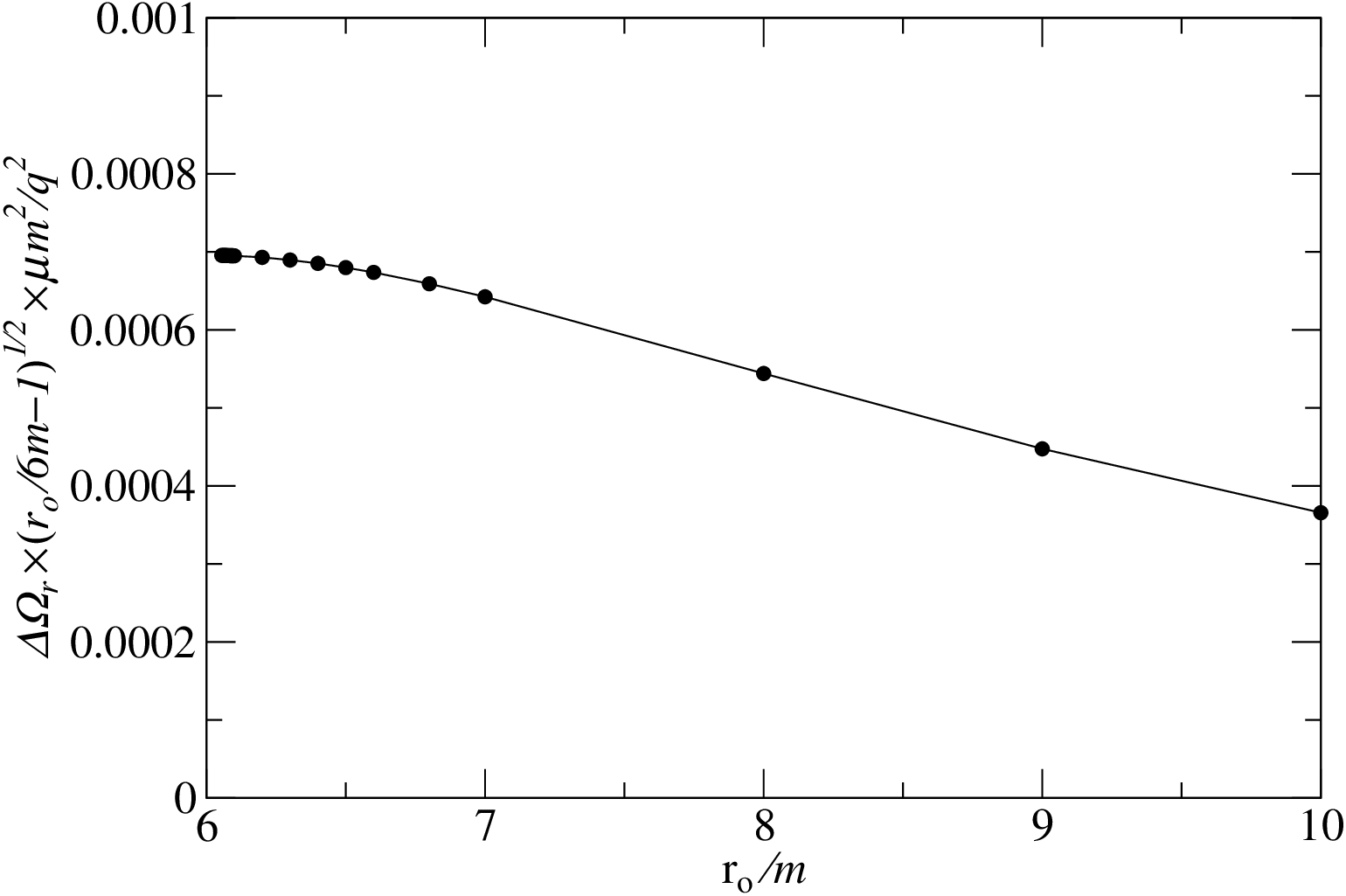}
 \caption{\label{FinOmra} For $q^2/\mu \ll \ro-6m $, the change in $\Omega_r$ from the
 self force as a function of the radius $\ro$ for slightly eccentric orbits
 close to the black hole. The limiting behavior at $q^2/\mu \ll \ro-6m <<
 m$ is given in \Deqn{DeltaOmrb}. See \Deqn{DeltaOmrc} for $q^2/\mu
 \approx \ro-6m$.}
 \end{figure}


To examine the effect of the self force on $\Omega_r$ for $\ro
\approx 6m$, let $f$ represent all but the first term on the right
hand side of \Deqn{Omegar2}. The change in $\Omega_r^2$ caused by the
self force is then
\beq
 \Delta(\Omega_r^2) = f = f_0 + f_1 (\ro-6m) + \cdots .
\eeq
Numerical analysis determines the values $f_0 = 9.46768\times10^{-5}
q^2/\mu m^3$, and $f_1 = -3.2318\times10^{-5} q^2/\mu m^2$.
 For $\ro=6m$,
\beq
  \Delta\Omega_r \underset{\ro\rightarrow6m} =
    f_0^{1/2} = 9.73020\times10^{-3} q/\mu^{1/2}m^{3/2} .
\Dlabel{DeltaOmrc}
\eeq

\begin{figure}
 \includegraphics[clip,angle=0,scale=0.32]{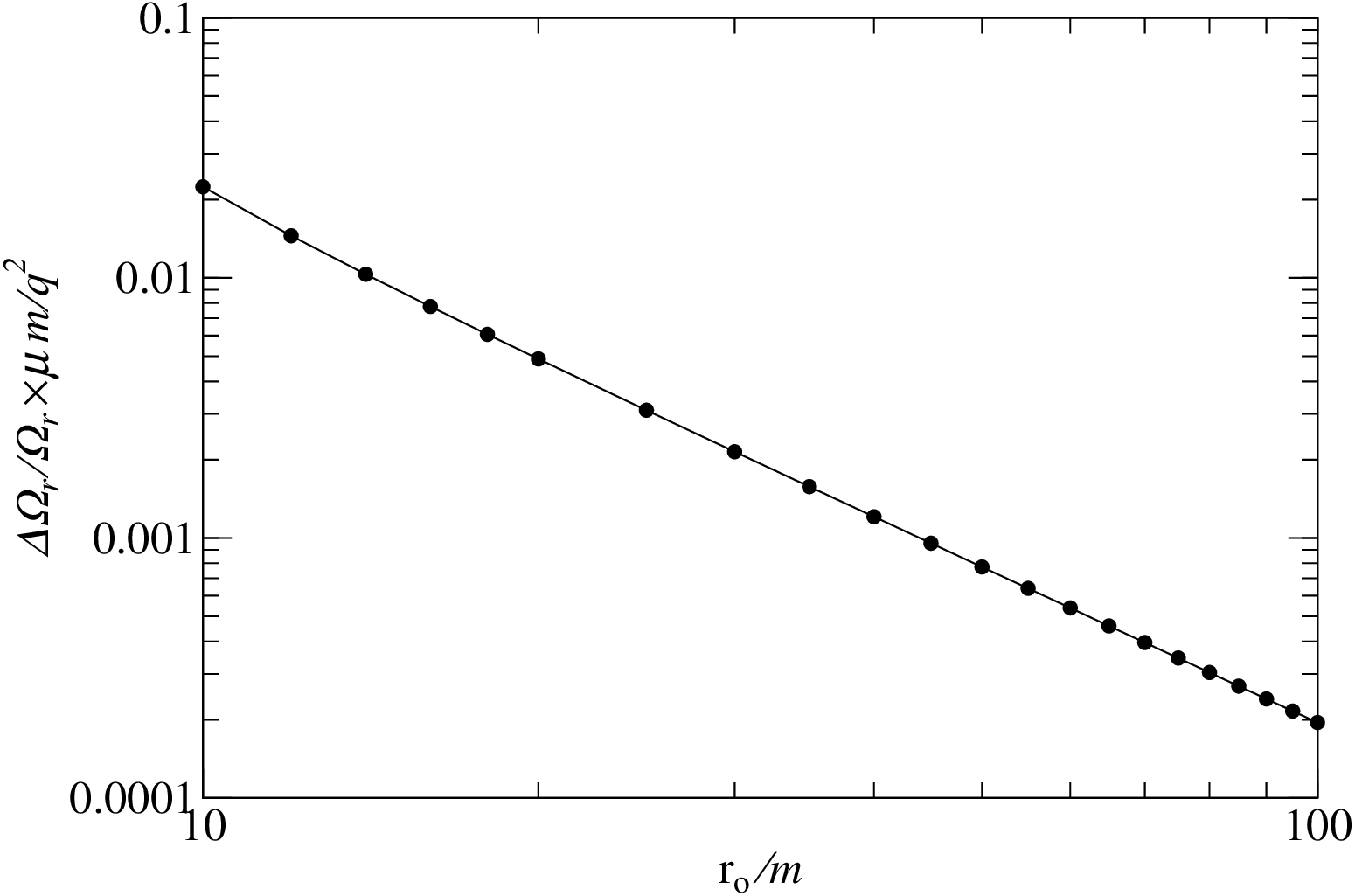}
 \caption{\label{FinOmrb} From numerical analysis, the fractional change in $\Omega_r$ from
 the self force as a function of the radius $\ro$ for slightly eccentric
 orbits far from the black hole.}
 \end{figure}

More generally, the change in $\Omega_r$ caused by the self force is
\bea
  \Delta\Omega_r
    &=& \left[\frac{m}{\ro^4}(\ro-6m) + f\right]^{1/2}
       - \left[\frac{m}{\ro^4}(\ro-6m)\right]^{1/2} .
\Dlabel{DeltaOmr}
\eea
When $\ro$ increases away from $6m$, $\Delta\Omega_r$ decreases and
changes scale
\bea
 \text{for } \ro-6m\gg q^2/\mu ,\quad
    \Delta\Omega_r &\approx& \frac{f\ro^2}{2} [m(\ro-6m)]^{-1/2}
\nonumber\\
      &=& O(q^2/\mu m^2).
\Dlabel{DeltaOmra}
\eea
The numerical calculation of $\Delta\Omega_r$, based upon
\Deqn{DeltaOmra}, is presented in Figs.~(\Dref{FinOmra}) and
(\Dref{FinOmrb}).
 Fig.~(\Dref{FinOmra}) illustrates the small end of this range, where
$q^2/\mu \ll \ro-6m \ll m$ and $f \approx f_0$. In this case,
\bea
  \Delta\Omega_r &\approx&
     6.95730\times10^{-4} (\ro/6m-1)^{-1/2} q^2/\mu  m^2
\Dlabel{DeltaOmrb}
\eea
gives the limit of the curve in Fig.~(\Dref{FinOmra}) as $\ro$ approaches
$6m$.

The right hand sides of \Deqns{DeltaOmrc} and (\Dref{DeltaOmrb}) are equal
at
\beq
  \frac{\ro}{6m} -1 \approx \frac{\ro^4}{24m^2} f_0
    = 0.00511255\, q^2/\mu m .
\eeq
At larger separations $\ro-6m \gtrsim m$,  $\Delta\Omega_r$ is still
approximated by \Deqn{DeltaOmra} and scales as $1/\ro^2$ as illustrated in
Fig.~(\Dref{FinOmrb}).

\subsection{Self force effects upon $\Omega_\pr$ for slightly
eccentric orbits}

The rate of precession of a slightly eccentric orbit
\beq
  \Omega_\pr \equiv \Omegao - \Omega_r
\eeq
is not particularly elegant when written in terms of $m$, $\ro$ and the
components of the self force.
 However,
\beq
 \Omega_\pr = \frac{m^{1/2}}{(6m)^{3/2}} - f_0^{1/2}
    \text{ for } \ro=6m .
\eeq
 In general,
\beq
  \Delta\Omega_\pr = \Delta\Omegao - \Delta\Omega_r,
\eeq
where $\Delta\Omegao$ and $\Delta\Omega_r$ may be obtained from
\Deqns{DeltaOm} and (\Dref{DeltaOmr}) .

\begin{figure}
\includegraphics[angle=0,scale=0.32]{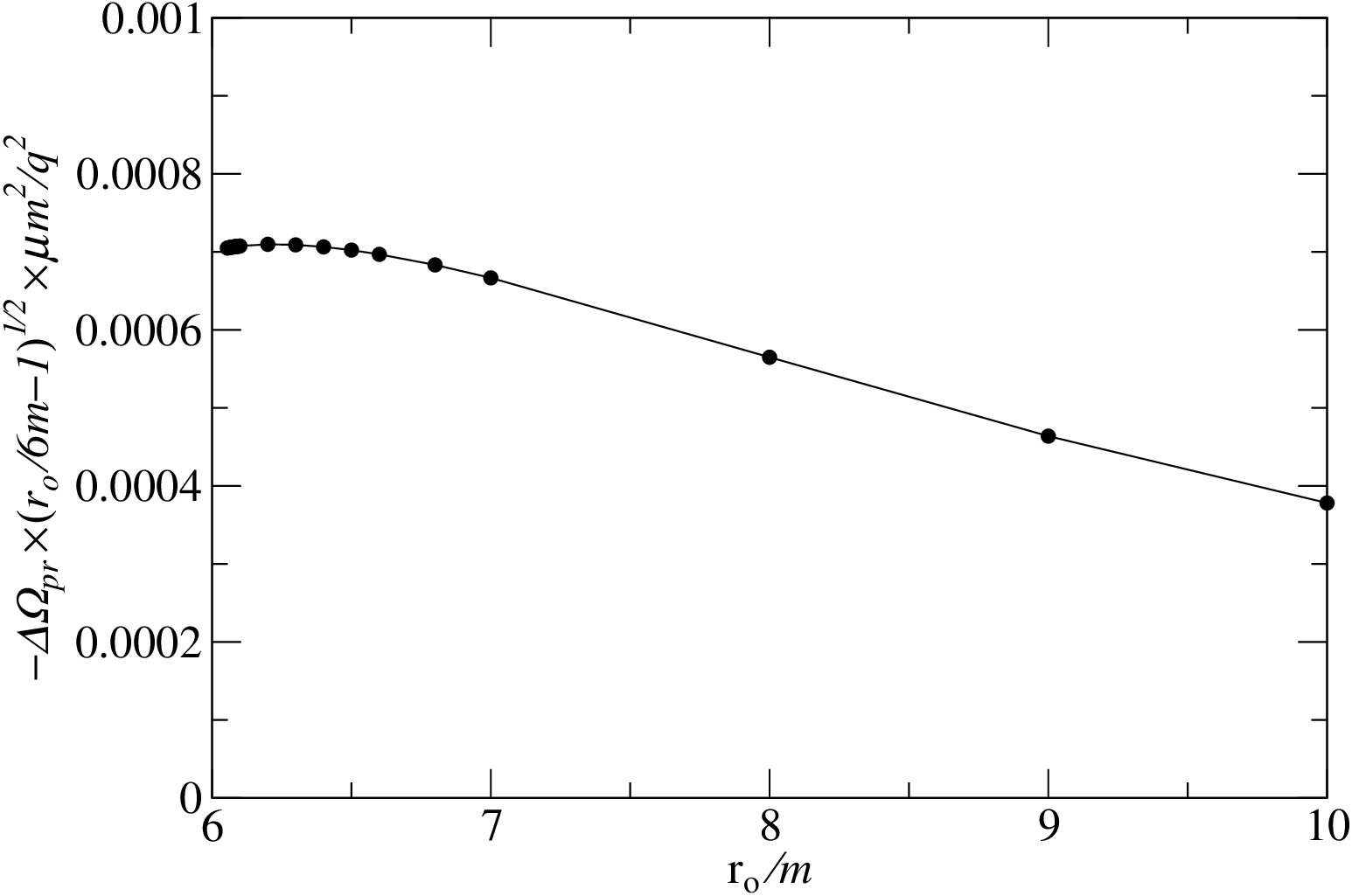}
\caption{\label{FinOmpra} For $q/m \ll \ro-6m$, the change in
$\Omega_\pr$, from the self force as a function of the radius $\ro$
for slightly eccentric orbits close to the black hole.}
 \end{figure}

\begin{figure}
\includegraphics[clip,angle=0,scale=0.32]{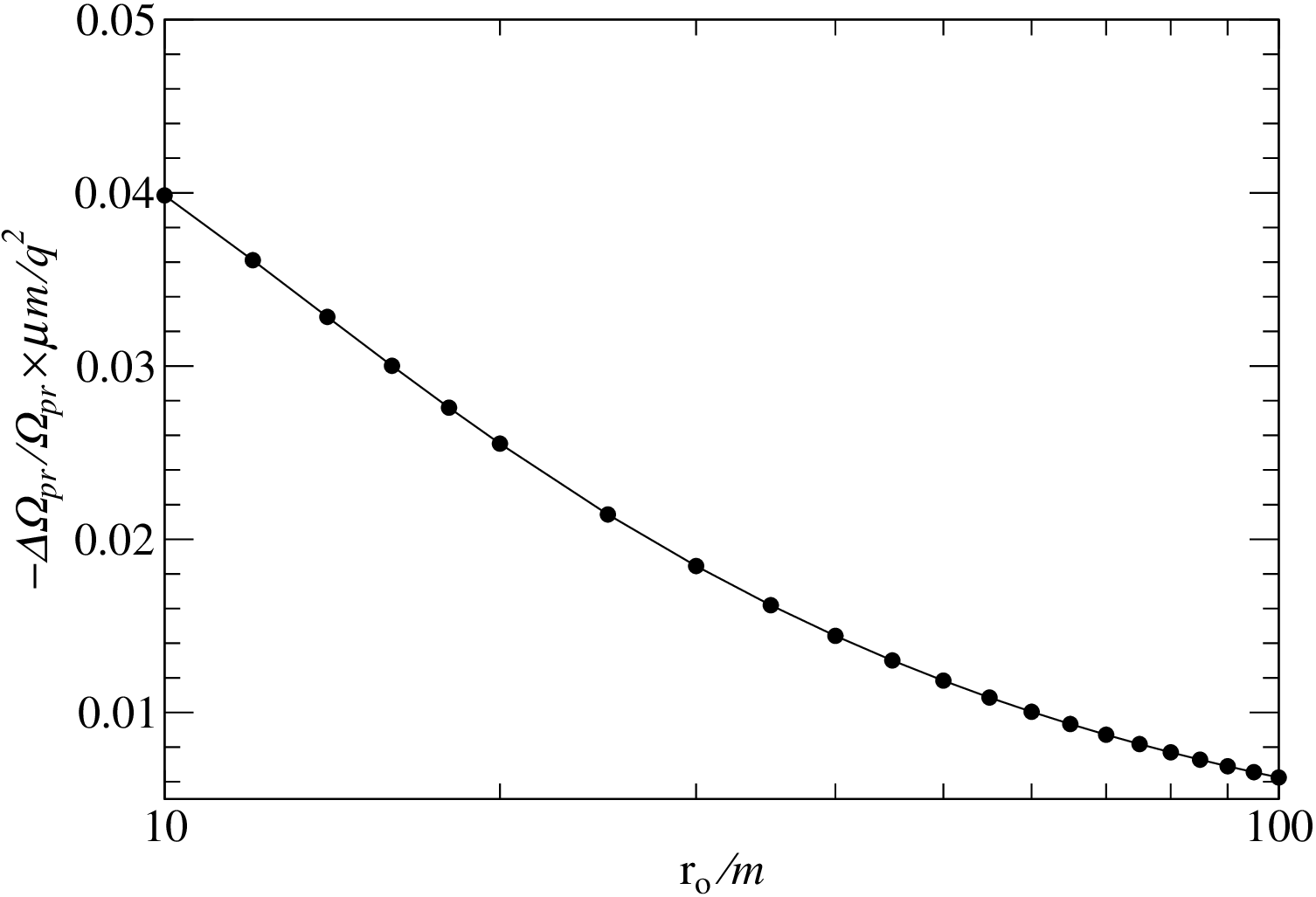}
\caption{\label{prec1b} For $q/m \ll \ro-6m$, the fractional change
in $\Omega_\pr$ from the self force as a function of the radius $\ro$
for slightly eccentric orbits far from the black hole.}
 \end{figure}

Fig.~\Dref{FinOmpra} illustrates $\Delta\Omega_\pr$ for orbits for
smaller values of $\ro$, and the curve for $\Delta\Omega_\pr$ has the
same limit to that of $-\Delta\Omega_r$ in Fig.~(\Dref{FinOmra}).
 The fractional change $\Delta\Omega_\pr/\Omega_\pr$ from the self force is
shown in Fig.~\Dref{prec1b} for larger values of $\ro$ where
$\Delta\Omega_\pr/\Omega_\pr$ scales as $\ro^{-1}$.

We summarize the falloff at large $\ro$ for a variety of quantities in
Table II.

Our earlier manuscript \Dcite{DetMessWhiting03} described the
numerical evaluation of $\partial_r\psi^\R_\o$ for a circular orbit
in great detail.
 The analytically known regularization parameters (the multipole moments of
$\partial_r\psi^\s$) were subtracted from the numerically determined
$\partial_r\psi_\o^\lm$. A few additional regularization parameters
were determined numerically and also subtracted from
$\partial_r\psi_\o^\lm$. The remainder was summed over $\ell$ up to
about 40.

The main difficulty revolved around the evaluation of
$\partial_r\psi_\o^\lm$ with sufficient  accuracy that the final sum
gave us good precision.
 Starting with approximately $13$ significant digits for
$\partial_r\psi_\o^\lm$, after the regularization parameters were
subtracted about $8$ significant digits remained. The main numerical
task for evaluating the scalar field self-force effects on $\Omega_r$
and $\Omega_\pr$ is very similar to this earlier work. A significant
difference, however, is the need to compute $F_t$, $F_\phi$ and
$F_r$, introduced in \Deqns{Ft}-(\Dref{Fr}). The details required for
determining the part of the field which depends upon the slight
eccentricity of the orbit are described in Appendix
\Dref{sourcedecomp}, and the regularization parameters are given in
Appendix \Dref{regularizationparameters}.

\section{Self force effect on the ISCO}
\Dlabel{isco}

The innermost stable circular orbit is characterized as that orbit for
which $\Omega_r$ is zero. The self force changes the radius of the ISCO
from $6m$ by
 $\Delta R$ where
\bea
  \Delta R_\isco &\equiv& (R_\isco - 6m)
\Dlabel{delRISCO}
\\
  &=& 180m^2 \partial_r\psio + (3/2) dJ^2/dR + 432 m^3 F_r .
\nonumber
\eea
 This result follows from equating the coefficient in square
brackets in \Deqn{Omegar} to zero, replacing $\ro$ by $6m$ in all terms
which are $O(q^2/\mu m)$ and solving for $(\ro-6m)$. The self-force
correction to the angular frequency of the ISCO is given by \Deqn{Omega2}
when evaluated at $\ro=6m+\Delta R_\isco$,
\bea
  \Omega^2_\isco &=& \ro^{-4}m(\ro-3\Delta R_\isco)
      - \ro^{-2}(\ro-3m) \partial_r \psio
\nonumber\\ &=&
   m(6m)^{-3} - 3m\Delta R_\isco (6m)^{-4}
\nonumber\\ && \qquad{}
      - (12m)^{-1} \partial_r \psio .
\Dlabel{OmISCO}
\eea
The fractional change in $\Omega_\isco$ from the scalar field self-force
is
\beq
  \Delta \Omega_\isco/ \Omega_\isco = - \Delta R_\isco/4m - 9 m \partial_r \psio .
\Dlabel{DeltaOmISCO}
\eeq

A summary of our numerical results for the effect of the scalar field
self-force on the ISCO is given in Table \Dref{ISCOtable}. Our numerical
work primarily followed that for evaluating $\Omega_\pr$. However, one
subtlety involved the need for evaluating both $F_t$ and $F_\phi$ in the
limit that $r\rightarrow6m$. For that task we found each quantity at
approximately thirty points between $r=6m$ and $7m$. A polynomial fit to
these data, using a variety of subsets of the thirty points and different
numbers of polynomial coefficients, provided robust results for the
required limits.


 \begin{table}
 \caption{ \label{scaling}
 Falloff at large $\ro$ for a variety of interesting quantities
 involving the scalar-field self-force.}
 \begin{tabular}{r l}
  Quantity & \quad behavior \\
   $\psi^\R_\o$ & $\sim\;\ro^{-3}$\\
   $\partial_r\psi^\R_\o$ & $\sim\;\ro^{-5}$\\

   $\Omegao$ & $\sim\;\ro^{-3/2}$   \\
   $\Delta\Omegao$  & $\sim\;\ro^{-9/2}$   \\

   $\Omega_r$ & $\sim\;\ro^{-3/2}$   \\
   $\Delta\Omega_r$ & $\sim\;\ro^{-7/2}$   \\

   $\Omega_\pr$ & $\sim\;\ro^{-5/2}$   \\
   $\Delta\Omega_\pr$ & $\sim\;\ro^{-7/2}$   \\

   $E$ & $\sim\;\ro^0$ \\
   $\Delta E$ & $\sim\;\ro^{-4}$ \\

   $J$ & $\sim\;\ro^{1/2}$ \\
   $\Delta J$ & $\sim\;\ro^{-5/2}$ \\
 \end{tabular}
 \end{table}

\begin{table}
\caption{\label{ISCOtable} Quantities of interest regarding the
 ISCO.}
\begin{ruledtabular}
\begin{tabular*}{3in}{ r @{\hspace{.2in}}  l }
 \hspace{.75in} $\Delta R_\isco \times \mu/q^2$       & --0.122701\hspace{.75in} \\
  $\Delta \Omega_\isco/ \Omega_\isco\times\mu m/q^2$ &  0.0291657 \\
\end{tabular*}
\end{ruledtabular}
\end{table}

\section{Discussion}
\Dlabel{discussion}

Two parts of our analysis of the self force effects on slightly eccentric
orbits are algebraically taxing while not particularly difficult
conceptually. These are the calculations of the regularization parameters
and the matching of the homogeneous solutions of the field equation across
the orbit of the particle.  While these two steps are individually
challenging, our confidence in the ultimate results is bolstered by the
appropriate convergence of the sum over the $\lm$ modes of the components
of the self force.  An error in either the analytical or the numerical
work involving either the matching or the regularization parameters would
be immediately heralded by a lack of convergence of the sum over modes.
Consequently, we deem these results trustworthy.

We see that the effect of the self force from a scalar field on the
innermost stable circular orbit of the Schwarzschild geometry is to move
the ISCO inward and to increase its angular frequency.  It is tempting to
generalize this result to the gravitational self force case.
 But we will not do so. There is a significant difference between the self
force effects of a scalar and a gravitational field.  In particular, as we
mentioned in the introduction, for gravity the self force already has an
important effect at the Newtonian level \Dcite{DetPoisson04}.  Namely it
is responsible for ensuring that the particle and the black hole both
orbit their common center of mass. This is not the case for the scalar
field.
 The difference may be traced back to the fact that a black hole has no
scalar charge. The charged particle responds only to the gravitational
interaction with the black hole and the scalar field interaction with its
own $\psi^\R$, and the particle motion deviates from a geodesic of the
Schwarzschild geometry only because of its scalar field. In this limit,
the black hole is not affected at all and remains fixed in space. And the
particle orbit is centered upon the black hole, not upon the common center
of mass.  To see the motion about the center of mass, it is necessary to
consider the gravitational self-force problem which we will return to in a
later paper in this series.

A well defined formulation of the stability of the ISCO requires the
imposition of equal amounts of outgoing and ingoing radiation at the
boundaries to make the system dissipation-free.
 However, actual slow inspiral into a black hole has only outgoing
radiation, and one might wonder about the relevance of our calculation to
the actual physical system.
 Ori and Thorne \Dcite{Ori00b} outline a careful treatment of the actual
slow evolution of a small object inspiralling  and making the transition
to a plunge into a black hole. They also indicate how self force effects
could be included in their analysis. An extension of our results in the
manner outlined by them is beyond the scope of this manuscript, but may be
returned to in the future.

\begin{acknowledgments}
Portions of this work were discussed at the sixth Capra meeting held in
June of 2003 at the Yukawa Institute for Theoretical Physics in Kyoto,
Japan, with additional support from Monbukagaku-sho Grant-in-Aid for
Scientific Research Nos. 14047212 and 14047214. We thank the organizers
(Norichika Sago, Misao Sasaki, Hideyuki Tagoshi, Takahiro Tanaka, Hiroyuki
Nakano, and Takashi Nakamura) for their hospitality during this meeting.
Portions of this work were further discussed at the seventh Capra meeting
held in June of 2004 at the Center for Gravitational Wave Astronomy,
University of Texas, Brownsville.  We thank the Center and Carlos Lousto
for hospitality during this meeting.  The participants of all of the
recent Capra meetings have provided many fruitful discussions.

This work was supported by the National Science Foundation under grant
PHY-0245024 with the University of Florida and under grant PHY-0200852
with the University of Wisconsin-Milwaukee.
\end{acknowledgments}

\appendix

\section{Source decomposition for nearly circular orbits}
\Dlabel{sourcedecomp}

In this Appendix $m$ refers to the spherical harmonic index while $M$
refers to the mass of the black hole. Elsewhere in this manuscript $m$ is
used for both of these quantities without confusion.

The effects of the self force on $\Omega_r$ and on the ISCO are
governed by the scalar field $\psi$ for a slightly eccentric
geodesic. Finding the field in this case is more difficult than for a
circular orbit. Apostolatos, et
al.\Dcite{ApostolatosKennefickOriPoisson93}, consider the
stress-energy for a point mass moving along such a geodesic, in their
Sections IV.B and IV.C, and provide an expansion of the source in
powers of a small constant $\dR$. Their analysis is easily modified
for a scalar charge.

A slightly eccentric geodesic in the equatorial plane is described by
\beq
  R(t) =  \ro + \dR \cos(\Omega_r t) 
\Dlabel{R(t)}
\eeq and
\beq
  \Phi(t) = \Omegao t + \delta\Phi(t)
          = \Omegao t + \frac{d \Omega_\phi}{dR}
               \frac{\dR}{\Omega_r} \sin (\Omega_r t),
\Dlabel{Phi(t)}
\eeq
where the angular frequency of the orbit is
\beq
  \Omega_\phi \equiv \frac{d\phi}{dt}
         = \Omegao + \frac{d\Omega_\phi}{dR} \dR \cos(\Omega_r t) .
\eeq
The frequencies $\Omegao$ and $\Omega_r$ are given in \Deqns{Kepler}
and (\Dref{Mercury}). The quantity $d\Omega_\phi/dR$ represents the
change in angular frequency with respect to a change in radius
\beq
  \frac{d\Omega_\phi}{dR} = -\frac{2(\ro-3M)}{\ro(\ro-2M)} \Omega_\phi,
\eeq
while the conserved quantities $E$ and $J$ are held constant at their
circular orbit values,
\beq
  E^2 = \frac{(\ro-2m)^2}{\ro(\ro-3m)}
         \qquad  J^2 = \frac{m\rotwo}{\ro-3m}.
\Dlabel{E2J2eqn}
\eeq

The scalar-field source which moves on such an orbit is
\bea
 \varrho &=& q \int_{-\infty}^\infty (-g)^{-1/2} \,
                  \delta[r-R(s)] \,
                  \delta[\phi-\Phi(s)]\,
\nonumber\\ &&
     {}\times \delta(\theta-\pi/2)\,
                  \delta[t-T(s)] \, ds
\nonumber\\ &=&
             \frac{q}{r^2} \left(\frac{dt}{ds}\right)^{-1}
     \,  \delta\!\left[r- \ro - \dR \cos(\Omega_r t)\right]
\nonumber\\ && \hspace{-.3in}
     {}\times  \delta\!\!\left[\phi- \Omegao t
  - \frac{d \Omega_\phi}{dR} \frac{\dR}{\Omega_r} \sin (\Omega_r t)\right]
       \delta(\theta-\pi/2) ,
\eea
after an integration over $s$ and substitutions from \Deqn{R(t)} and
(\Dref{Phi(t)}). An expansion of the delta functions about the orbit,
with $\dR$ being small, along with a spherical harmonic
decomposition of this source, which includes an integration by parts
over the angle $\phi$, yields
\bea
 \varrho_\lm(t,r)
    &=& \oint
       \varrho Y^\ast_\lm(\theta,\phi) \,d\Omega
\nonumber\\ &=&
  \frac{q(\ro-2M)}{\ro^3 E}Y_\lm^\ast(\pi/2,0) e^{-im\Omega_\phi t}
\nonumber\\ &&
 {} \times
   \left[ 1 - im\frac{d\Omega_\phi}{dR} \frac{\dR}{\Omega_r}
   \sin(\Omega_r t)\right]
\Dlabel{sourceSpH}
\\ &&
 {}\times
   \left[\delta(r- \ro) - \dR\cos(\Omega_r t)\delta^\prime(r- \ro)\right] .
\nonumber
\eea

This reveals that the source has a frequency spectrum consisting of
the harmonics of the angular frequency $\omega_m \equiv m \Omegao$
along with sidebands at frequencies $\omega_m^\pm \equiv \omega_m \pm
\Omega_r$. The amplitude of the sidebands are proportional to $\dR$.

The scalar field wave equation is
\beq
  \nabla^2\Psi = -4\pi\varrho .
\Dlabel{del2psi}
\eeq
\begin{widetext}
The separation of variables of $\Psi$ yields
\beq
  \Psi = \sum_{\lm} \Psi_\lm(t,r,\theta,\phi)
    = \sum_{\lm \omega} \Psi_{\lm}^\omega (r) e^{-i\omega t} Y_\lm(\theta,\phi),
\eeq
where $\ell=0\ldots\infty$, $m=-\ell\ldots \ell$ and $\omega =
\{\omega_m, \omega_m^-,\omega_m^+\} $.
 The radial equation for the $r$ dependence of any $\ell$ mode
with a frequency $\omega$ is
\beq 
  \frac{d^2 \Psi^\omega_\lm(r)}{dr^2}
       + \frac{2(r-M)}{r(r-2M)}  \frac{d\Psi^\omega_\lm(r)}{dr}
       + \left[\frac{\omega^2 r^2}{(r-2M)^2} - \frac{\ell(\ell+1)}{r(r-2M)} \right]
            \Psi^\omega_\lm(r)
  = - \frac{ 4\pi \varrho^\omega_\lm}{1-2M/r} .
\Dlabel{diffeq}
\eeq

For self force calculations it is convenient to use $\psi\equiv
(q/\mu)\Psi$ and to divide each $\psi_\lm$ into the $\omega_m$ part
$\psi^\o_\lm$ and the side band parts $\dR\chi^\pm_\lm$,
\beq
  \psi_\lm(t,r,\theta,\phi) = \left( \psi^\o_\lm e^{-i\omega_m t}
    + \dR \chi^-_\lm e^{-i\omega^-_m t}
    + \dR \chi^+_\lm e^{-i\omega^+_m t}\right) Y_\lm(\theta,\phi).
\eeq
\end{widetext}
The numerical determination of these parts of $\psi_\lm$ requires the
source-free solutions of \Deqn{diffeq} with appropriate boundary
conditions, and then the proper match of these solutions across the
orbit of the particle at $\ro$. The matching conditions, from
\Deqns{sourceSpH} and (\Dref{diffeq}), are
\bea
  \left[\psi_\lm^\o\right]_\ro &\equiv& \lim_{\epsilon\rightarrow0^+}
      \left[\psi_\lm^\o(\ro+\epsilon)
      - \psi_\lm^\o(\ro-\epsilon) \right]
\Dlabel{jumppsio}
\nonumber\\  &=& 0
\\
  \left[\frac{d\psi_\lm^\o}{dr}\right]_\ro &=&
             - \frac{4\pi q^2}{\mu\ro^2 E} Y^*_\lm(\pi/2,0)
\Dlabel{jumpdpsio}
\eea
\bea
  \left[\chi_\lm^\pm\right]_\ro &=&
              \frac{2\pi q^2}{\mu\ro^2 E} Y^*_\lm(\pi/2,0)
\Dlabel{jumpchi}
\\
\left[\frac{d\chi_\lm^\pm}{dr}\right]_\ro
     &=& - \frac{4\pi q^2}{\mu\ro^2 E} Y^*_\lm(\pi/2,0)
\nonumber\\ &&
       \times \left[ \frac{M}{\ro(\ro-2M)}
                    \pm \frac{m}{2\Omega_r} \frac{d\Omega_\phi}{dR}
                       \right]
\Dlabel{jumpdchi}
\eea
where, in the Appendix only, $[\ ]_\ro$ on the left hand side denotes
the discontinuous change in a quantity across the orbit at $\ro$.

\begin{widetext}

The scalar field $\psi$ is a purely real field, and it is convenient
to combine the $m$ and $-m$ contributions
\bea
  \psi_\lm  + \psi_\lmm
     &=& (\psi^\o_\lm + \psi^{\o*}_\lm) \cos[m(\phi-\Omegao t)]
            Y_\lm(\theta,0)
   + (\chi^+_\lm + \chi^{+*}_\lm) \dR \cos[m(\phi-\Omegao t) - \Omega_r t]
            Y_\lm(\theta,0)
\nonumber\\ &&{}
   + (\chi^-_\lm + \chi^{-*}_\lm) \dR \cos[m(\phi-\Omegao t) + \Omega_r t]
            Y_\lm(\theta,0),
\nonumber\\ &&{}
\Dlabel{psilm+psilmm}
\eea
where we have used the fact that $\psi$ is the conservative field
given by $\frac12(\psi^\ret + \psi^\adv)$. The contribution from
$m=0$ for a given $\ell$ is just one half of that in
\Deqn{psilm+psilmm} with $m$ set to zero,
\bea
  \psi_{\ell0}
     &=& \frac12 (\psi^\o_{\ell0} + \psi^{\o*}_{\ell0})
            Y_{\ell0}(\theta,\phi)
   + (\chi^+_{\ell0} + \chi^{-}_{\ell0}) \dR \cos(\Omega_r t)
            Y_{\ell0}(\theta,\phi).
\Dlabel{psil0}
\eea
Note that $\chi^-_{\ell0} = \chi^{+*}_{\ell0}$ for the conservative
field.

In Section \Dref{isco} we require $\psi_\lm+\psi_\lmm$ and its
derivatives evaluated at the particle.  The location of the particle
is given in \Deqn{R(t)} and (\Dref{Phi(t)}). For $m \ne 0$, an
expansion about $\ro$, retaining only terms through first order in
$\dR$, yields
\bea
  \left(\psi_\lm + \psi_\lmm \right)_p
     &=& \left(1 + \dR(t)\frac{\partial}{\partial r}
                 + \delta\Phi(t)\frac{\partial}{\partial \phi} \right)
     \times\left<(\psi^\o_\lm + \psi^{\o*}_\lm) \cos[m(\phi-\Omegao t)]
            Y_\lm \right>
\nonumber\\ &&{}
   + (\chi^+_\lm + \chi^{+*}_\lm + \chi^-_\lm + \chi^{-*}_\lm)
     \dR \cos(\Omega_r t) Y_\lm,
\eea
where the subscript $p$ implies evaluation at the particle. From here
through the remainder of this Appendix \Dref{sourcedecomp}, the spherical
harmonic $Y_\lm$ is to be evaluated at $(\pi/2,0)$, and the other terms on
the right hand sides are to be evaluated at $(r, \phi) =(\ro, \Omegao t)$
only after all appropriate derivatives have been taken. After
simplification, this becomes
\bea
  \left(\psi_\lm + \psi_\lmm \right)_p
     &&= (\psi^\o_\lm + \psi^{\o*}_\lm) Y_\lm
    + \big( \partial_r\psi^\o_\lm + \partial_r \psi^{\o*}_\lm
    + \chi^+_\lm + \chi^{+*}_\lm + \chi^-_\lm + \chi^{-*}_\lm \big)
   \dR \cos(\Omega_r t) Y_\lm.
\Dlabel{psilm+-m}
\eea

Similar expansions start with \Deqn{psilm+psilmm} and ultimately  provide
the $\phi$ derivative
\bea
  \frac{\partial}{\partial \phi} \left(\psi_\lm + \psi_\lmm \right)_p
     &=& \bigg[-\frac{m^2}{\Omega_r} \frac{d\Omega_\phi}{dR}
       (\psi^\o_\lm + \psi^{\o*}_\lm)
        +  m (\chi^+_\lm + \chi^{+*}_\lm
                   - \chi^-_\lm - \chi^{-*}_\lm) \bigg]
     \dR \sin(\Omega_r t) Y_\lm,
\Dlabel{dpsilmdphi+-m}
\eea
the $t$ derivative
\bea
  \frac{\partial}{\partial t} \left(\psi_\lm + \psi_\lmm \right)_p
     &=& \bigg[\frac{m^2 \Omegao}{\Omega_r} \frac{d\Omega_\phi}{dR}
       (\psi^\o_\lm + \psi^{\o*}_\lm)
        - \omega^+_m (\chi^+_\lm + \chi^{+*}_\lm)
        + \omega^-_m (\chi^-_\lm + \chi^{-*}_\lm) \bigg]
    \dR \sin(\Omega_r t) Y_\lm,
\Dlabel{dpsilmdt+-m}
\eea
and the $r$ derivative
\bea
  \left(\partial_r\psi_\lm + \partial_r\psi_\lmm \right)_p
     &&= (\partial_r\psi^\o_\lm + \partial_r\psi^{\o*}_\lm) Y_\lm
        + \Big(\partial^2_r \psi^\o_\lm
                     + \partial^2_r  \psi^{\o*}_\lm
\nonumber\\ &&{} \quad
   + \partial_r\chi^+_\lm + \partial_r\chi^{+*}_\lm
   + \partial_r\chi^-_\lm + \partial_r\chi^{-*}_\lm \Big)
     \dR \cos(\Omega_r t) Y_\lm.
\Dlabel{dpsilmdr+-m}
\eea
\end{widetext}

In Section \Dref{isco} we require the sum over $\ell$ and $m$ of
these three previous derivatives. Accordingly, we define $F_t$,
$F_\phi$ and $F_r$ from
\beq
  (\partial_t\psi^\R)_p = - F_t \Omega_r \dR \sin(\Omega_r t)
\Dlabel{defFt}
\eeq
\beq
  (\partial_\phi\psi^\R)_p =
          - F_\phi \Omega_r \dR \sin(\Omega_r t)
\Dlabel{defFphi}
\eeq
and
\beq
  (\partial_r\psi^\R)_p = (\partial_r\psio)_\ro
   + F_r \dR \cos(\Omega_r t),
\Dlabel{defFr}
\eeq
where the required regularization is described in Appendix
\Dref{regularizationparameters}.

\section{Regularization parameters for $\psi$}
\Dlabel{regularizationparameters}

\newcommand{\p}{\prime}
\newcommand{\Epsi}{\epsilon}
\newcommand{\rhoB}{\tilde{\rho}}

We describe the regularization of the scalar field as developed by
Barack and Ori \Dcite{BarackOri00, BMNOS02} for a particle in a
circular orbit of the Schwarzschild geometry. Our notation follows
that of Refs.~\Dcite{DetMessWhiting03, Messaritaki03, DongHoon04}.
All of these are required reading for a thorough understanding of
this Appendix.

The scalar field is regularized at the location of the particle by
subtracting the singular part of the field $\psi^\s$ from the actual
field $\psi$. The remainder $\psi^\R \equiv \psi - \psi^\s$ is then
guaranteed \Dcite{DetWhiting03} to be a regular solution of the
vacuum scalar field equation in the vicinity of the particle, and the
derivatives of $\psi^\R$ at the particle provide the required
self-force resulting from the particle interacting with its own
field.

The mode sum regularization procedure \Dcite{BarackOri00, BMNOS02}
describes the multipole decomposition of $\psi^\R$ in terms of the
decompositions of $\psi$ and $\psi^\s$,
\bea
 \psi^\R &=& \sum_{\lm} \psi_\lm^\R(t,r) Y_\lm(\theta,\phi)
\nonumber\\ &=&
   \sum_{\lm} \left[\psi_{\lm}(t,r) - \psi^\s_{\lm}(t,r) \right]
   Y_\lm(\theta,\phi).
\Dlabel{psiretlm}
\eea

The numerical determination of $\psi_{\lm}(t,r)$ for a slightly
eccentric orbit is discussed in Appendix \Dref{sourcedecomp} and in
Ref.~\cite{DetMessWhiting03}. $\psi^\s$, however, is singular at the
location of the particle, and only well defined in a neighborhood of
the particle. Nevertheless, its multipole decomposition over a
two-sphere of radius $r$ is finite, even if $r$ coincides with the
the radial coordinate of the particle. The decomposition is not
unique because of the ambiguity in the definition of $\psi^\s$ away
from the particle. However, the mode sum regularization procedure
remains valid because its sum is only required in a neighborhood of
the particle, where the sum must equal $\psi^\s$. Thus, in evaluating
$\psi^\R$ and its derivatives at the particle the individual
$\psi^\s_{\lm}$ in \Deqn{psiretlm} are not unique but the sum in
\Deqn{psiretlm} and its derivatives converge to unique values.

Barack and Ori \Dcite{BarackOri00, BMNOS02} find it convenient to do
the sum over $m$ first and then to describe the multipole
decomposition of a derivative of $\psi^\s$ as
\beq
 (\partial_a \psi^\s)_p = \sum_{\ell}
      \left[\left( \ell+\frac12 \right)A_a + B_a
          + \frac{C_a}{\ell+\frac12} + O(\ell^{-2}) \right],
\Dlabel{Fabco}
\eeq
where the $O(\ell^{-2})$ terms yield precisely zero when summed over
$\ell$. The constants $A_a$, $B_a$ and  $C_a$ are independent of
$\ell$, and are determined by a multipole decomposition of an
expansion of $\psi^\s$ about the location of the particle.

The required regularization parameters for the derivatives of
$\partial_t\psi$ are derived from Eqs.~(8a)-(8d) of
Ref.~\Dcite{BMNOS02}. We discuss only those parameters which have not
previously appeared in an actual application\Dcite{Burko00,
DetMessWhiting03}. In our notation, with $\dot{R}$ representing a
derivative of $R$ with respect to proper time $s$, these
regularization parameters are
\beq
  A_{\pm t} = \pm \frac{ q^2 \dot{R}}{\mu(R^2+J^2)}
\eeq
\beq
  A_{\pm r} = \mp \frac{q^2E}{\mu R^2 (E^2 - \dot{R}^2)}
\eeq
\beq
  A_{\pm \phi} = 0
\eeq
\beq
  B_{t} = \frac{q^2E R\dot{R}}{2\mu(R^2+J^2)^{3/2}}
                       \left(F_{1/2}- 2F_{-1/2}\right)
\eeq
\beq
  B_{r} = - \frac{q^2R^2
    \left[
      (2E^2-\dot{R}^2)F_{1/2} - (E^2+\dot{R}^2) F_{-1/2}
     \right]}{2\mu(R-2M)(R^2+J^2)^{3/2}}
\eeq
\beq
  B_{\phi} = \frac{q^2 R \dot{R}(F_{1/2} - F_{-1/2})}
                                   {2 \mu J (R^2+J^2)^{1/2}}
\eeq
\beq
  C_t = C_r = C_\phi = 0 .
\eeq
Here the hypergeometric function is represented by
 $F_q \equiv {}_2F_1(q,\frac12;1;z)$ where the argument $z = M/(R-2M)$.

These parameters may be expanded by use of \Deqn{R(t)}, which implies that
\beq
  \dot{R} = - \frac{\dR E\Omega_r \sin(\Omega_r t)}{1-2M/R}.
\eeq
Through first order in $\dR$ the non-zero regularization parameters
for a slightly eccentric orbit are
\bea
  A_{\pm t} &=&
    \mp \frac{q^2\Omega_r}{\mu\ro^2}
        \dR \sin(\Omega_r t)
\eea
\beq
  A_{\pm r} = \mp \frac{q^2}{\mu\ro^2 E}
     \left[1 -\frac{2}{\ro} \dR\cos(\Omega_r t)  \right]
\eeq
\bea
  B_{t} &=&
   - \frac{q^2\Omega_r\left(F_{1/2}- 2F_{-1/2}\right)}{2\mu\ro(\ro^2+J^2)^{1/2}}
     \dR \sin(\Omega_r t)
\eea
\bea
  B_{r} &=&
   - \frac{q^2 \left( 2 F_{1/2} - F_{-1/2} \right)}
                {2\mu \ro (\ro^2+J^2)^{1/2}}
\nonumber\\ && {} \times
   \left[1 - \frac{2\ro^2+J^2}{\ro(\ro^2+J^2)}
                     \dR\cos(\Omega_r t)\right]
\nonumber\\ && {}
    + \frac{q^2 M\left( 2 F^\prime_{1/2} - F^\prime_{-1/2} \right)}
                {2\mu \ro(\ro-2M)^2 (\ro^2+J^2)^{1/2}}
         \dR\cos(\Omega_r t)
\nonumber\\ && {}
\eea
\bea
  B_{\phi} &=&
  - \frac{q^2 \Omega_r (F_{1/2} - F_{-1/2})}{2 \mu J (1-2M/\ro)^{1/2}}
          \dR \sin(\Omega_r t)
\eea
In this expansion, $F^\prime_{1/2}$ is the derivative of the
hypergeometric function $F_{1/2}$ with respect to its argument $z$. Both
$F_{1/2}$ and $F^\prime_{1/2}$ are evaluated at $z=M/(\ro-2M)$.

The regularization parameters for $F_t$ and $F_\phi$, defined in
\Deqns{defFt} and (\Dref{defFphi}), are obtained by removing the
factor $-\Omega_r\dR\sin(\Omega_r t)$ from $A_{\pm t}$, $B_t$ and
$B_\phi$. Similarly, the regularization parameters for $F_r$, defined
in \Deqn{defFr}, are obtained by removing the factor
$\dR\cos(\Omega_r t)$ from the $\dR$ terms of $A_{\pm r}$ and $B_r$.

The regularization parameters for the scalar field, alone, warrants
further discussion \Dcite{Messaritaki03}. In a particular locally-inertial
coordinate system $(T,X,Y,Z)$, the singular field near a scalar charged
particle is simply
\beq
  \psi^\s = q/\rho + O(\rho^3/\calR^4)
\eeq
where $\calR$ is a length scale of the geometry in the vicinity of the
particle, and $\rho^2 = X^2 + Y^2 + Z^2$. For the special case that the
particle is in a circular orbit about a Schwarzschild black hole, a
coordinate transformation between the special $(T,X,Y,Z)$ coordinates and
the usual Schwarzschild coordinates allows $\rho$ to be written as a
function of Schwarzschild coordinates, and the expansion of $1/\rho$ about
$\rho=0$ is given in Eq.~(6.22) of Messaritaki \Dcite{Messaritaki03}. The
terms of interest are
\begin{equation}
\frac{1}{\rho} =  \epsilon^{-1} \frac{1}{\tilde{\rho}}
 + \epsilon^{1} \left[ \frac{\ro-3m}{8 \rotwo (\ro-2m)}
        \left( \frac{1}{\chi} -
                 \frac{(\ro + m)}{\ro} \frac{1}{\chi^2} \right) \tilde{\rho}
          \right] {} + \ldots
\label{OrderRhoa}
\end{equation}
where $\ldots$ refers to terms which vanish as $r\rightarrow0$ in such a
manner that they have no contribution to the regularization parameters. In
this equation we use
\beq
 \rhoB^2 \equiv \frac{\ro(r-\ro)^2}{\ro-2m}
          + 2 \rotwo \frac{\ro-2m}{\ro-3m} \chi (1-\cos\Theta)
\label{rhoBdef}
\eeq
where
\beq
 \chi \equiv 1 - \frac{m\sin^2\Phi}{\ro - 2m}.
\label{chidef}
\eeq
The angles $(\Theta,\Phi)$ are derived from a rotation of the
Schwarzschild coordinates which puts the particle on the $\Theta=0$ axis
\Dcite{DetMessWhiting03}.

We find that, through the $\Epsi^1$ term of \Deqn{OrderRhoa}
\beq
  \psi^\s_{\ell0}(r=\ro) = B_\psi - \frac{2\sqrt{2} D_\psi}{(2\ell-1)(2\ell+3)}
\eeq
where $B_\psi$ and $D_\psi$ result from the $\Epsi^{-1}$ and $\Epsi^{1}$
terms of \Deqn{OrderRhoa} respectively. Appendices C and D of
Ref.~\Dcite{DetMessWhiting03} describe the expansion of the $\Theta$
dependence in terms of Legendre polynomials and a convenient method of
finding the $m=0$ component by integrating over the angle $\Phi$.

For $\Delta=0$, the $\epsilon^{-1}$ term is
\beq
  \frac{1}{\tilde\rho} = \sqrt{\frac{(\ro-3m)}{2 \rotwo(\ro-2m)}}
     \chi^{-1/2} (1-\cos\Theta)^{-1/2} .
\eeq
From Eqs.~(C3) and (D7) in \cite{DetMessWhiting03}, this gives
\beq
 B_\psi = \sqrt{\frac{(\ro-3m)}{2 \rotwo(\ro-2m)}} \, F_{1/2} \sqrt{2}.
\eeq

For $\Delta=0$, the $\epsilon^{1}$ term is
\begin{widetext}
\beq
     \frac14 \sqrt{\frac{(\ro-3m)}{2 \rotwo (\ro-2m)}}
               \left[\frac{1}{\chi} - \frac{(\ro+m)}{\ro\chi^2} \right]
               \chi^{1/2}
     (1-\cos\Theta)^{1/2}.
\eeq
From Eqs.~(C3) and (D16) in \cite{DetMessWhiting03} this gives
\beq
   D_\psi\frac{-2\sqrt{2}}{(2\ell-1)(2\ell+3)} =
        \frac14 \sqrt{\frac{(\ro-3m)}{2 \rotwo (\ro-2m)}}
               \left[F_{1/2} - \frac{(\ro+m)}{\ro} F_{3/2} \right]
               \frac{-2\sqrt{2}}{(2\ell-1)(2\ell+3)} .
\eeq
\end{widetext}

\bibliographystyle{prsty}

\end{document}